\documentclass[%
 reprint,
superscriptaddress,
 amsmath,amssymb,
 aps,
 longbibliography,
]{revtex4-1}

\usepackage{float}
\usepackage{epsfig}
\usepackage{amsmath}
\usepackage{amsfonts}
\usepackage{amssymb}
\usepackage{comment}
\usepackage{array}
\usepackage{tabularx}
\usepackage[version=3]{mhchem} 
\usepackage{color}
\usepackage{siunitx}
\usepackage[version=3]{mhchem}
\usepackage{xcolor}
\usepackage{diffcoeff}
\usepackage[normalem]{ulem}

\newcommand{\cartoon}[1]{\begin{minipage}[c]{.04\textwidth}
      \includegraphics[width=6mm, height=6mm]{#1}
       \end{minipage}}

\newcolumntype{C}{>{\centering\arraybackslash}p{2.5cm}}
\newcolumntype{L}{>{\centering\arraybackslash}m{3cm}}
\newcolumntype{P}[1]{>{\RaggedRight\hspace{0pt}}p{#1}}
\newcolumntype{K}[1]{>{\centering\arraybackslash}p{#1}}

\newcommand{\fig}{\text{Fig. }}
\newcommand{\sect}{\text{Sec. }}
\newcommand{\eqn}{\text{Eqn. }}
\newcommand{\ES}{\text{Supplementary Material }\cite{SI}\text{ }}
\newcommand{\apn}[1]{\text{Appendix } \ref{sec:A#1}}
\DeclareSIUnit\Molar{\textsc{m}}
\begin{document}

\title{The Symmetry Basis of Pattern Formation in Reaction-Diffusion Networks}

\author{\text{Ian Hunter}}
\thanks{These two authors contributed equally}
\affiliation{
Brandeis University Physics, Waltham MA, 02453 USA
}%

\author{Michael M. Norton}
\thanks{These two authors contributed equally}
\affiliation{%
Center for Neural Engineering, Department of Engineering Science and Mechanics, The Pennsylvania State University, University Park, PA 16802 USA
}%

\author{Bolun Chen}
\affiliation{
Brandeis University Volen National Center for
Complex Systems, Waltham MA, 02453 USA}
\affiliation{
 Department of Physics Boston University, Boston MA, 02215 USA
}%
\author{\text{Chris Simonetti}}
\affiliation{
Brandeis University Physics, Waltham MA, 02453 USA
}
\author{\text{Maria Eleni Moustaka}}
\affiliation{
Brandeis University Physics, Waltham MA, 02453 USA
}%
\author{Jonathan Touboul}
\affiliation{
Brandeis University Volen National Center for
Complex Systems, Waltham MA, 02453 USA}
\affiliation{
Brandeis University Mathematics Department}

\author{\text{ Seth Fraden}}
\email[To whom correspondence should be addressed:]{fraden@brandeis.edu}
\affiliation{
Brandeis University Physics, Waltham MA, 02453 USA 
}%

\begin{abstract}

In networks of nonlinear oscillators, symmetries place hard constraints on the system that can be exploited to predict universal dynamical features and steady-states, providing a rare generic organizing principle for far-from-equilibrium systems. However, the robustness of this class of theories to symmetry-disrupting imperfections is untested. Here, we develop a model experimental reaction-diffusion network of chemical oscillators to test  applications of this theory in the context of self-organizing systems relevant to biology and soft robotics. The network is a ring of 4 identical microreactors containing the oscillatory Belousov-Zhabotinsky reaction coupled to nearest neighbors via diffusion. Assuming perfect symmetry, theory predicts 4 categories of stable spatiotemporal phase-locked periodic states  and 4 categories of invariant manifolds that guide and structure transitions between phase-locked states. In our experiments, we observed the predicted symmetry-derived synchronous clustered transients that occur when the dynamical trajectories coincide with invariant manifolds. However, we observe only 3 of the 4 phase-locked states that are predicted for the idealized homogeneous system.  Quantitative agreement between experiment and numerical simulations is found by accounting for the small amount of experimentally determined heterogeneity.  This work demonstrates that a surprising degree of the network’s dynamics are constrained by symmetry in spite of the breakdown of the assumption of homogeneity and raises the question of why heterogeneity destabilizes some symmetry predicted states, but not others.
\end{abstract}
\date{\today}


\maketitle

\section{Introduction}

Network science unifies the study of disparate physical systems that can be cast as discrete sets of interacting dynamical units \cite{Strogatz2001}. Here, we focus on networks of self-driven oscillators for which this simple framework provides profound insights into systems ranging from electrical power grids to biological neural networks known as central pattern generators (CPG) responsible for coordinating autonomous animal locomotion \cite{Motter2013,Golubitsky1999,Takamatsu2001,In2003,Matheny2019}.

\par 
The design of networks that generate bespoke spatiotemporal patterns is a great challenge because universal organizing principles for far-from-equilibrium systems are exceedingly rare. Exploiting network symmetry is one way to meet this challenge. Symmetries place hard constraints on the network dynamics of self-driven oscillators by dictating that certain transient features and steady-state patterns \emph{must} exist.  Specifically, the theory of equivariant dynamics describes how the symmetries of the network affect the symmetry of the network dynamics\cite{Stewart2015}. For a given symmetry operation, such as a permutation of network nodes, some sets of points in state-space  will be unchanged and consequently, their dynamics must also remain the same \cite{Ashwin1992,Stewart2015}.

\par A class of results derived from group theory arises by combining the spatial symmetry of the network with the temporal symmetry of the oscillators, providing a natural framework for describing spatiotemporal patterns\cite{Ashwin1992,Golubitsky1999,Golubitsky2000,Golubitsky2006,Stewart2015,Golubitsky2016}. One, the H/K theorem, allows enumeration of all symmetry derived patterns in which phase-locked nodes co-evolve because they receive the same input from their neighbors\cite{Golubitsky2000}. Remarkably, some of the predicted patterns are far from obvious and bear little resemblance to the geometric symmetry of the network. Significantly, these patterns are universal. They depend only on the coupling topology and are independent of all system specific details regarding the nature of the non-linear oscillators themselves and even whether or not the coupling is non-linear. However, these striking results  derive from the strong assumption that classes of nodes in the network and their interconnections are strictly identical \cite{Ashwin1992,Golubitsky2016}

Golubitsky and colleagues applied the H/K theorem, along with a few plausible assumptions, to make a surprising prediction in neuroscience; the minimal network architecture of central pattern generators in all quadrupeds can be determined simply by cataloguing the aggregate of gaits observed across species. Or, in other words, that form follows function in neuronal networks\cite{Golubitsky1999}. The existence of such CPGs is controversial in the case of mammals, but evidence exists for other organisms \cite{Couzin-Fuchs2015,zhang_neural_2014-3,Kopell1988}.

In this work, we experimentally study oscillatory  chemical reaction-diffusion networks and examine the dynamics through the lens of symmetry-based network theories\cite{Ashwin1992,GolubitskyFull2000}. The significance of studying a self-contained reaction-diffusion system lies in the potential for fabrication of autonomous devices that organize their spatiotemporal dynamics through processes analogous to living systems.

Our goal here is to  ascertain whether symmetry can serve as a conceptual basis and engineering principle for the structuring of spatiotemporal patterns on chemical networks, which can be equally applied to understanding biological neural networks and engineering chemical networks for soft robotics\cite{Litschel2018}.  

The title of this paper is a paean to Turing who was the first to consider the theory of pattern formation in discrete reaction-diffusion networks \cite{Turing1952}. Although the majority of Turing's paper, ``The chemical basis of morphogenesis,'' focused on static, spatially varying patterns, Turing also predicted spatiotemporal pattern formation, including standing and traveling chemical waves, which have been observed in chemical networks\cite{tompkins_testing_2014}. It was this latter aspect of Turing's theory that motivated us to experimentally test symmetry-based network theory using coupled chemical oscillators. However, because Turing's linear stability analysis is limited to the onset of pattern formation, we were motivated to employ the symmetry approach of the H/K theory because it is universal, holding true for non-linear oscillators and for all time.

\par  To test the network theory, we develop a minimally complex   experimental system consisting of a ring of 4 identical, nanoliter sized, chemical reactors containing the Belousov-Zhabotinsky (BZ) oscillating reaction and coupled to nearest neighbors by diffusion. This experimental system oscillates stably for about 70 periods\cite{Sheehy2020}, which is an order of magnitude longer than reported in previous studies of self-organized networks\cite{Takamatsu2001,Takamatsu2006,Tayar2017}. To thoroughly explore state-space, we perform hundreds of trials  by running experiments simultaneously on multiple copies of the network resulting in an order of magnitude greater number of experiments than done previously with different chemical networks\cite{Litschel2018}. We model this reaction-diffusion network at two levels of description. The most detailed is a mathematical model of the system explicitly describing the BZ reaction chemistry, which we assume occurs only in the reactors, with coupling between nearest neighbors caused by diffusion of a subset of the BZ chemicals through the intervening PDMS. We theoretically reduce this reaction-diffusion network into a simpler phase model to analyze the predictions of the theory of equivariant dynamics. These models contain far fewer free parameters than independent measurements. This, the large ensemble of experiments and their longevity allows quantitative comparisons between theory and experiment leading to firm conclusions regarding the applicability of idealized network dynamics to model the steady-state and transient dynamics of this self-organized system in which both the oscillators and coupling are fully chemical.

Our experiments reveal an intricate array of transient and phase-locked spatiotemporal chemical dynamics. However, when we reduce these complex non-stationary solutions into the state-space of phase relationships, the H/K theorem allows us to represent high-dimensional chemical dynamics in terms of simple, model-independent geometric objects in the form of planes, lines, and points that are readily visualized. This geometric perspective leads to an appreciation of aspects of the experimental behaviors that are directly imposed by symmetries, thereby providing conceptual understanding of the complex dynamics, which complements and enriches the quantitative comparison between theory and experiment. 

\begin{figure}
\includegraphics[scale=.1]{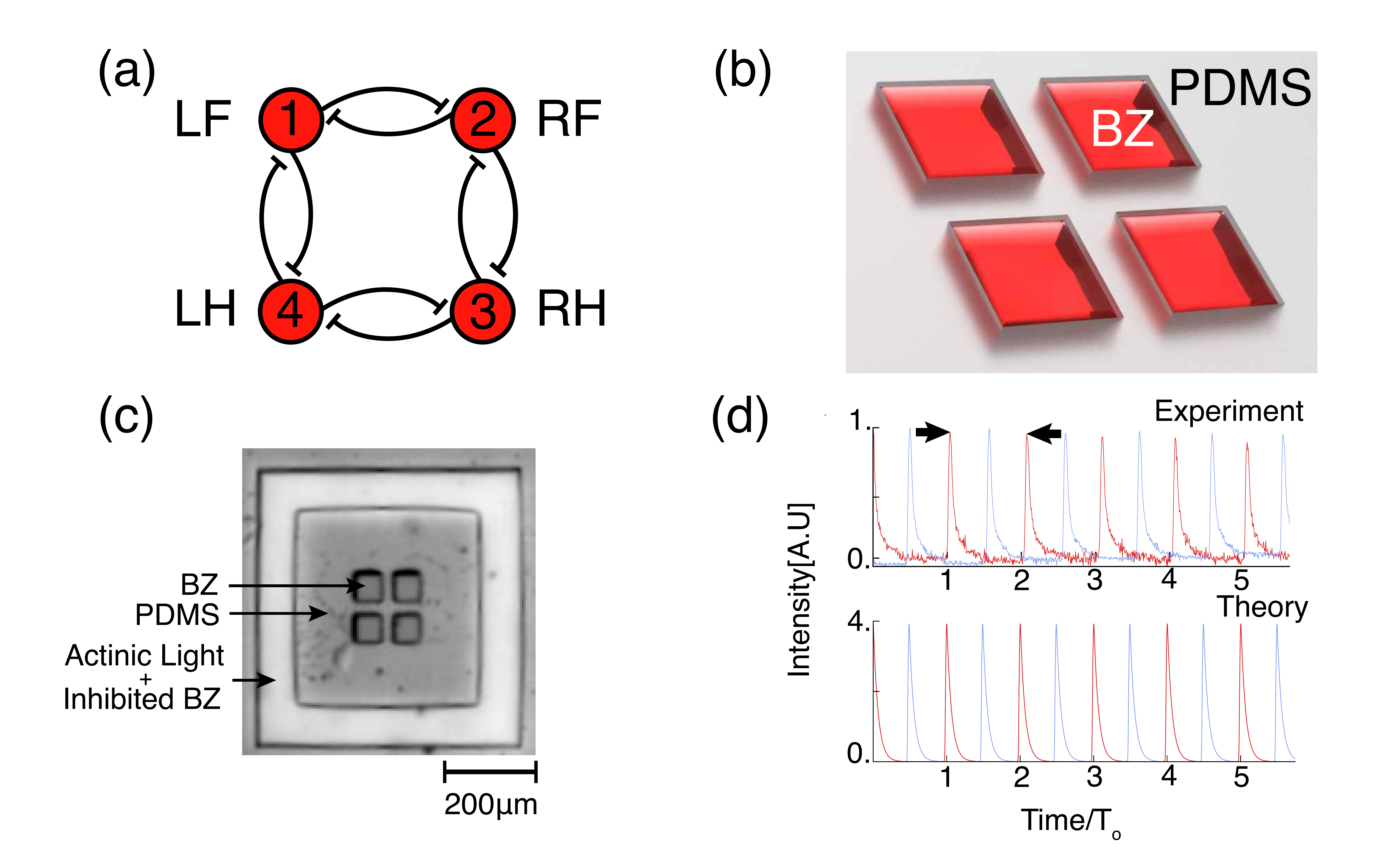}
 \caption{ \textbf{(a)}   Schematic of a network of a ring of 4 inhibitory coupled oscillators.  Indexing of nodes is indicated as either a number (1,2,3,4), or leg of a quadruped (LF, RF, RH, LH) with L left, R right, F front and H hind.  \textbf{(b)} Schematic of the experimental system. The reactors are divots in the PDMS, filled with BZ and sealed between 2 glass plates.  \textbf{(c)} Photograph of BZ filled 4-ring network. Actinic light illuminates BZ in a channel surrounding the network and provides a constant chemical boundary condition. \textbf{(d)}  Two adjacent reactors (red and blue traces) in the network oscillating $180^\circ$ out-of-phase with each other. \textbf{Top:} Measured transmitted intensity versus time. \textbf{Bottom:} Simulated oxidized catalyst concentration [mM] versus time. In both, time is rescaled by oscillation period $T_0$, indicated by the two arrows.
 }
  \label{fig:introdiagram}
\end{figure}

\section{Results}
\subsection{Experimental Reaction-Diffusion Network}
\par
We designed a  reaction-diffusion network consisting of a ring of four diffusively coupled nanoliter volume batch reactors laid out in a square 2x2 lattice with nearest neighbor coupling (see \fig \ref{fig:introdiagram}). Previously, we employed emulsions containing the BZ oscillating reaction to study reaction-diffusion networks\cite{Toiya2010,Delgado2011,li_combined_2014,tompkins_testing_2014,Tompkins2015,Li2015,Wang2016,Norton2019}. But the diffusive coupling between surfactant stabilized emulsion drops was difficult to characterize and manufacturing of the networks was challenging, which contributed to a large degree of variability between experiments.  Here, to improve reproducibility we manufactured these reactors to high precision from  elastomeric PDMS using soft lithography techniques and filled the reactors with the oscillatory BZ reaction as described previously\cite{Litschel2018,Sheehy2020}, illustrated in \fig \ref{fig:introdiagram} and in \apn{A}. To obtain a large statistical sample of trajectories we made devices that combined 9 or 16 copies of the 2x2 network \cite{Litschel2018}. To optimize homogeneity in the chemical concentrations of each of the reactors, we simultaneously filled the entire set of networks by pipetting a drop of BZ that floods all the reactors before sealing the sets of reactors by clamping the PDMS between two glass plates [\ES \fig 1-2, SI videos in \cite{Litschel2018}].

The chemical coupling between adjacent reactors arises from the permeation of chemical species through the intervening PDMS wall and mainly consists of bromine-induced inhibition, with a weaker activator coupling, perhaps  by bromous acid and the bromine dioxide radical \cite{li_combined_2014,Norton2019,Li2015,Litschel2018,vanag_model_2009,tompkins_testing_2014,Torbensen2017a,Wang2016,Proskurkin2018,Torbensen2017,Delgado2011,Vanag2011,Toiya2010}. After  mixing the BZ reagents, pipetting them onto the PDMS networks, sealing the networks,  and placing the sample in the dark for an induction period of 20 minutes, it was observed that all reactors began to oscillate and collectively form spatiotemporal patterns [\fig \ref{fig:introdiagram}(d)] \cite{Litschel2018}. 

 The reactors form a closed system and consequently  the oscillators have a finite lifetime as the reactants are consumed and waste products accumulate. However, although the amplitude of the chemical oscillations decreases over time, the oscillators  maintain a nearly constant period for a duration of order 70 oscillations\cite{Sheehy2020}. Based on this long term stability, we assume that the underlying phase dynamics of the individual BZ oscillators remains constant during the duration of the experiment, thus allowing us to study phase relationships between reactors as they evolve over time [\ES  \fig 4, Movies S1-4]. Each 4-ring network is isolated from the environment because the reactors are surrounded by a zone of photosensitive BZ that is held at constant chemical conditions by the application of actinic light\cite{Litschel2018}. We also assume that each chemical reactor is well mixed,   ignoring any  spatial variation of chemical concentrations, because the size of the reactor is small compared with the length scale of diffusion,  e.g. $w < \sqrt{D \tau}$ with $w$ the width of each square reactor  ($w = 62 ~$\si{\micro \metre}), $D$, the diffusion constant of each BZ chemical ($D \sim 10^{-9}\mathrm{m}^2\mathrm{s}^{-1}$) and $\tau$, the duration of a BZ oscillation ($\tau \sim 300$s).

\par 

\subsection{Theory and the Role of Symmetry}

\par The fullest description  of the dynamics of the 4-ring network that we consider is a reaction-diffusion network model. It focuses on the time dependent concentrations of the well mixed chemicals in each  reactor, denoted as $\left( \bar{c}_1(t),\bar{c}_2(t),\bar{c}_3(t),\bar{c}_4(t) \right)$, where $\bar{c}_j(t)$ is a vector of concentrations in the $j_{
\mathrm{th}}$ reactor with indices as in \fig \ref{fig:introdiagram}A. Assuming that the reactors are identical, the behaviors are expected to have the same symmetries as a square, e.g. 3 rotations and 4 reflections. Associated with this symmetry group, the H/K theorem predicts invariant manifolds,  subspaces in which the dynamics remains confined, that are universal to any ring of 4 oscillators, enumerated in Table \ref{table:lock}. Although the theory is more general, we restrict ourselves to the case in which all the nodes are on the same limit cycle, as this corresponds to experiment. With this assumption the H/K theorem guarantees any system of 4 oscillators with square symmetry possesses 8 categories of invariant manifolds, including 4 categories of phase-locked periodic states. These states are therefore efficiently described by considering the phase relationship between pairs of reactors, defined as the fraction of period they are shifted from each other on their common limit cycle.

Invariant manifolds are denoted by a pair of symmetry operations, (H,K). The first symmetry, H, represents an exchange of nodes that results in the same state subject to one or more phase-shifts. The second symmetry, K, indicates symmetries under which the system is unchanged.

Depending on the constraints imposed by H and K, these solutions may either maintain fixed phase relationships among all four nodes resulting in phase-locked solutions, or leave 1- or 2-dimensional freedom on these relationships, as enumerated in Table \ref{table:lock}.
The 4 categories of phase-locked periodic states are spatiotemporal periodic patterns  that correspond to 6 point invariant manifolds in the phase difference space that can be identified with gaits of quadrupeds enumerated in Table \ref{table:lock} and visualized in \fig \ref{fig:libstates}(b).  The first two categories are \textit{Pronk} in which all the legs advance simultaneously and  \textit{Trot} for which diagonal legs are in phase, and the two diagonal pairs of legs are half a period out of phase.  \textit{Pace} and \textit{Bound} form one category and we refer to them interchangeably in the remainder of the text. In Pace, legs on each side are in phase and opposite sides out-of-phase, while for Bound, legs on opposite sides are in phase and the front legs out-of-phase with the hind legs.  \textit{Clockwise (counter clockwise) Rotary Gallop} is another category in which the legs advance in a clockwise (counter clockwise) manner with each leg advancing a quarter of a period later than the preceding leg. 

\par The remaining 4 categories correspond to higher dimensional invariant manifolds (lines or planes) that contain trajectories maintaining partial symmetries. Along 1-dimensional linear invariant manifolds, the network can be split into two pairs of reactors, such that within pairs the reactors are in phase or antiphase, while between pairs reactors have an arbitrary phase-shift. Along 2-dimensional manifolds, two nodes oscillate in phase and the other two nodes are at arbitrary phase-shifts. In fact, the 2-dimensional manifolds intersect  the 1-dimensional manifolds, and the 1D manifolds intersect the phase-locked 0-dimensional manifolds [\fig \ref{fig:libstates}(a)]. Heuristically, these higher dimensional manifolds often act as privileged pathways that both guide and structure transient transitions between the phase-locked states. Beyond predicting the existence of these invariants, the H/K theorem neither prescribes their stability nor precludes the existence of others. To address questions of stability and existence of additional manifolds requires a specific model of the oscillators and their connections.

\subsection{Chemical Kinetic and Phase Models}
\par To model the reaction-diffusion dynamics of our experiments we use the Vanag-Epstein model of the BZ reaction, which treats the chemical kinetics of 4 BZ chemicals, \ce{Br-}, \ce{HBrO_2}, Ferroin, and \ce{Br_2}, combined with a diffusive coupling term between chemical species in which the coupling strength is fitted from the data.  Noting that $\bar{c}_i\in \mathbb{R}^4$ are the concentrations of these 4 chemicals in reactor  $i\in \{1,2,3,4\}$ and $R_0:\mathbb{R}^4\mapsto \mathbb{R}^4$ denotes the Vanag-Epstein BZ model vector field~\cite{li_combined_2014,Norton2019,Li2015,Litschel2018,vanag_model_2009,tompkins_testing_2014,Torbensen2017a,Wang2016,Proskurkin2018,Torbensen2017,Delgado2011,Vanag2011,Toiya2010}, we obtain the equation: 

\begin{equation}
 \frac{d }{dt}\bar{c}_i = \bar{R}_0(\bar{c}_i) + \sum_{j=1}^4  A_{ij} \mu(\bar{c}_j-\bar{c}_i)
 \label{eqn:RDeqhom}
\end{equation}
where $\mu \in \mathbb{R}^{4\times 4}$ accounts for the chemical coupling matrix between two adjacent cells, and depends on the permeability of the PDMS for each chemical species and the geometry of the reactors, while $A\in \mathbb{R}^{4\times 4}$ denotes the adjacency matrix between two reactors that is determined by the network topology  [\apn{C} for details on the model, choice of parameters and fits of free parameters].  This model ignores spatial concentration gradients inside the reactors, effectively treating reactors as points corresponding to nodes of the network. The model also neglects the occurrence of chemical reactions within the PDMS, which acts as a connector that couples adjacent nodes.

\begin{center}
\begin{table}[h!]
\centering
\caption{Symmetry required invariant manifolds for an oscillator network possessing square or Dihedral 4 ($D_4$) symmetry.   $D_4$, all symmetries of a square; $D_n^p$, reflection across $n$ diagonals; $D_n^s$, reflection across $n$, vertical or horizontal, axes;  $Z_4$, $90^{\circ}$ rotation; $Z_2$, $180^{\circ}$ rotation;  $1$, no operation. The first 4 classes of manifolds are phase-locked states. The column marked ``Phase'' graphically indicates the spatiotemporal pattern with symbols representing the phase in percentage of the period $T_0$, white circle - $0\%$; white/black - $25\%$; black circle - $50\%$;   black/white - $75\%$.  $T_0$ denotes the period of each oscillator in a given invariant manifold and can vary from manifold to manifold. The second 4 classes of manifolds are symmetrically clustered states, related by arbitrary phase shifts $f_1,f_2$, which vary from 0 to 1 as fraction of a period. The graphical representation of nodes in the column ``Phase''  have solid, striped, or dot motifs. Different motifs are related by an arbitrary phase shift. Similar motifs with opposite background colors are antiphase with each other.}

\setlength{\tabcolsep}{0pt} 
\renewcommand{\arraystretch}{1} 
\begin{tabular}{ K{14mm} K{6.5mm} K{6mm} K{23mm} K{17mm} K{23mm} } 

\hline
\hline
\multicolumn{6}{c}{\text{\small Point invariant manifolds:} } \\
\hline

 \textbf{\tiny{Name}} \textbf{ \tiny{(H,K)}} & \textbf{\tiny{Phase}} & $\bar{c}_1$ & $\bar{c}_2$ & $\bar{c}_3$ & $\bar{c}_4$ \\ 
\hline
  \vspace{-4mm}  \text{\tiny{Pronk}} ${\scriptstyle (D_4,D_4)}$ &
    \cartoon{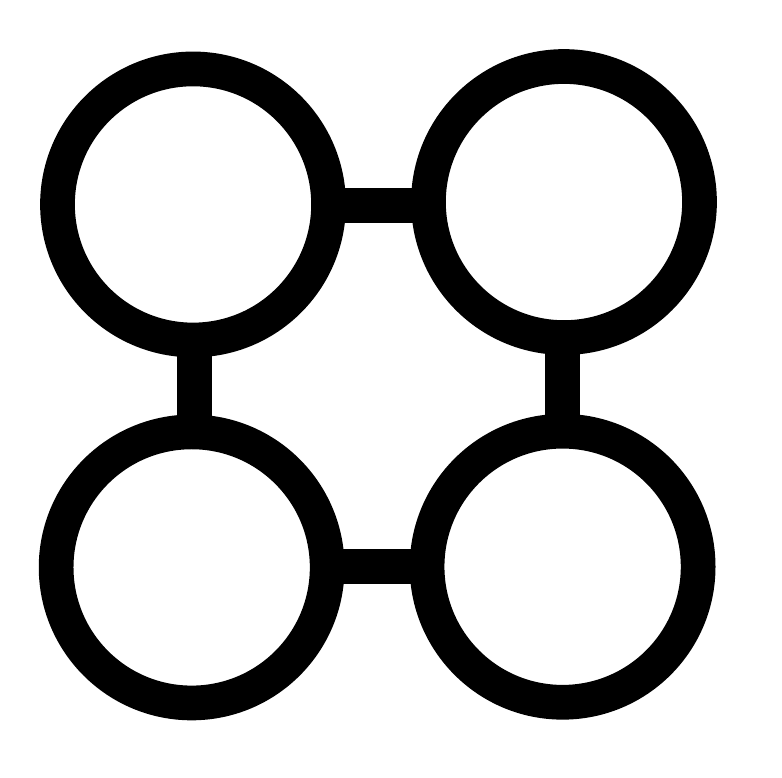} &
    $\bar{c}(t)$ & $\bar{c}(t)$ & $\bar{c}(t)$ & $\bar{c}(t)$ \\ 
\hline
      \vspace{-4mm} \text{\tiny{Trot}} ${\scriptstyle (D_4,D^p_2)}$&
    \cartoon{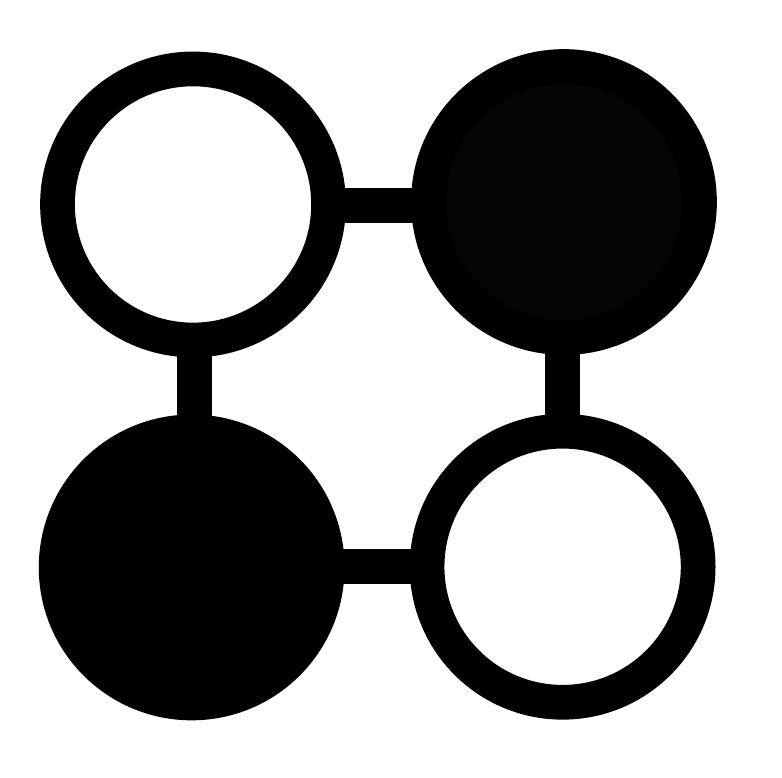} &
    $\bar{c}(t)$ & $\bar{c}(t+\frac{T_0}{2})$ & $\bar{c}(t)$ &  $\bar{c}(t+\frac{T_0}{2})$  \\ 
\hline
     \vspace{-4mm}  \par\text{\tiny Pace}  ${\scriptstyle (D^s_2,D^s_1)}_A$\par\text{\tiny Bound}  ${\scriptstyle (D^s_2,D^s_1)}_B$& 
    \cartoon{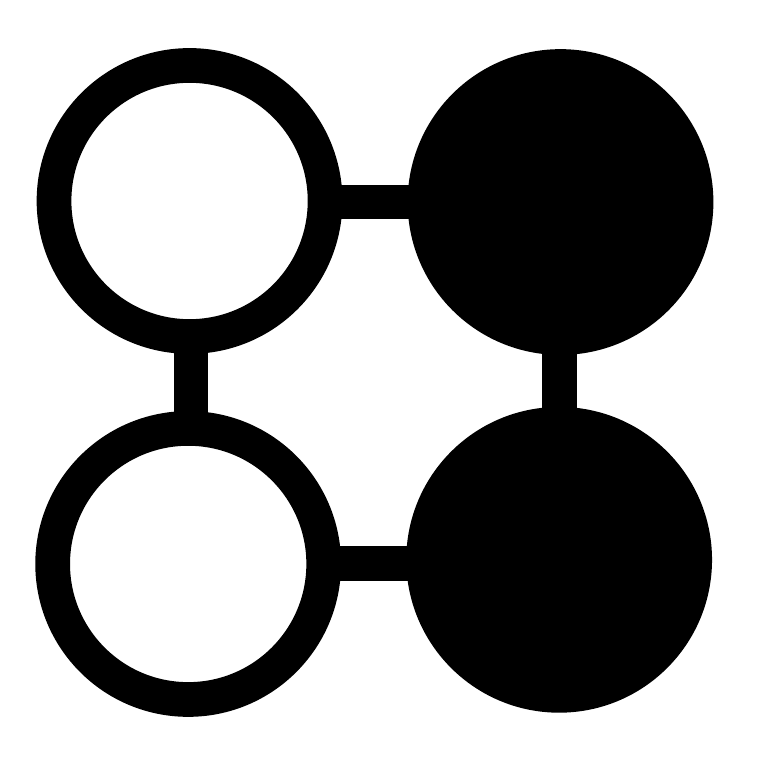}\newline\cartoon{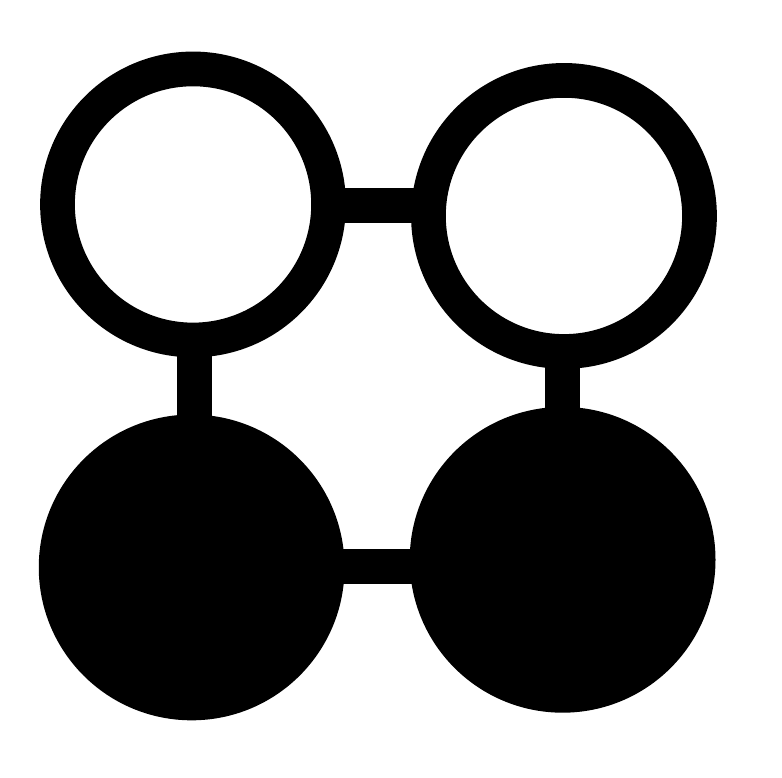} & 
    \par $\bar{c}(t)$ \par $\bar{c}(t)$ & 
    \par $\bar{c}(t+\frac{T_0}{2})$ \par $\bar{c}(t)$ & 
    \par $\bar{c}(t+\frac{T_0}{2})$ \par $\bar{c}(t+\frac{T_0}{2})$ & 
    \par $\bar{c}(t)$ \par $\bar{c}(t+\frac{T_0}{2})$   \\ 
\hline
   \vspace{-4mm}    \par\text{\tiny{CW Gallop}} 
     ${\scriptstyle (Z_4,1)}_A$
    \par\text{\tiny{ CCW Gallop}}
     ${\scriptstyle (Z_4,1)}_B$&
    \cartoon{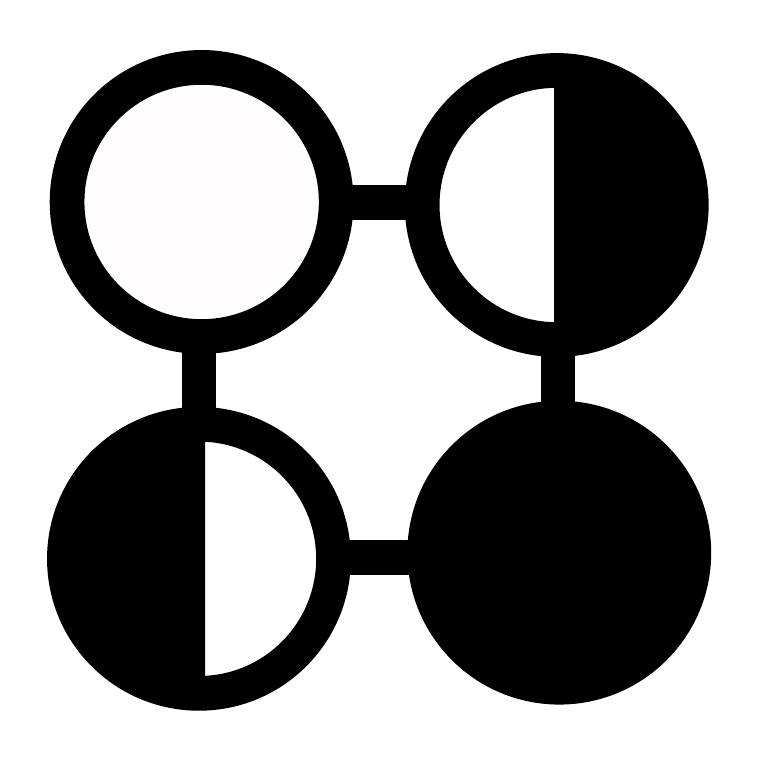}\newline\cartoon{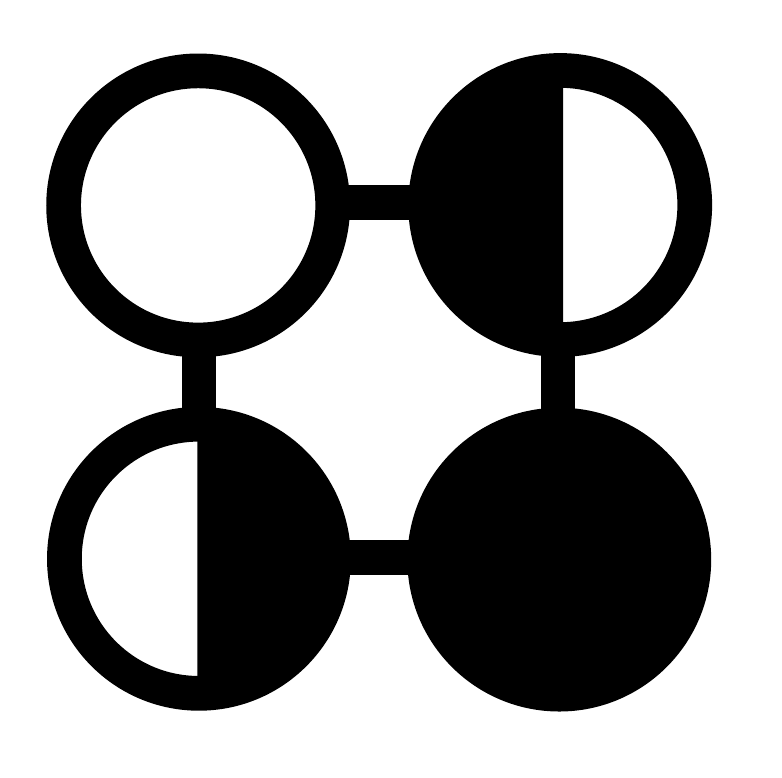}&
    \par$\bar{c}(t)$\par$\bar{c}(t)$ &
    \par$\bar{c}(t + \frac{T_0}{4})$\par$\bar{c}(t - \frac{T_0}{4})$ & \par$\bar{c}(t+\frac{T_0}{2})$\par$\bar{c}(t+\frac{T_0}{2})$ & 
    \par$\bar{c}(t - \frac{T_0}{4} )$\par$\bar{c}(t + \frac{T_0}{4} )$ \\ 
\hline
\multicolumn{6}{c}{\text{\small Linear invariant manifolds: }} \\
\hline

    ${\scriptstyle (D^s_1,D^s_1)}_A$ \newline${\scriptstyle (D^s_1,D^s_1)}_B$& \cartoon{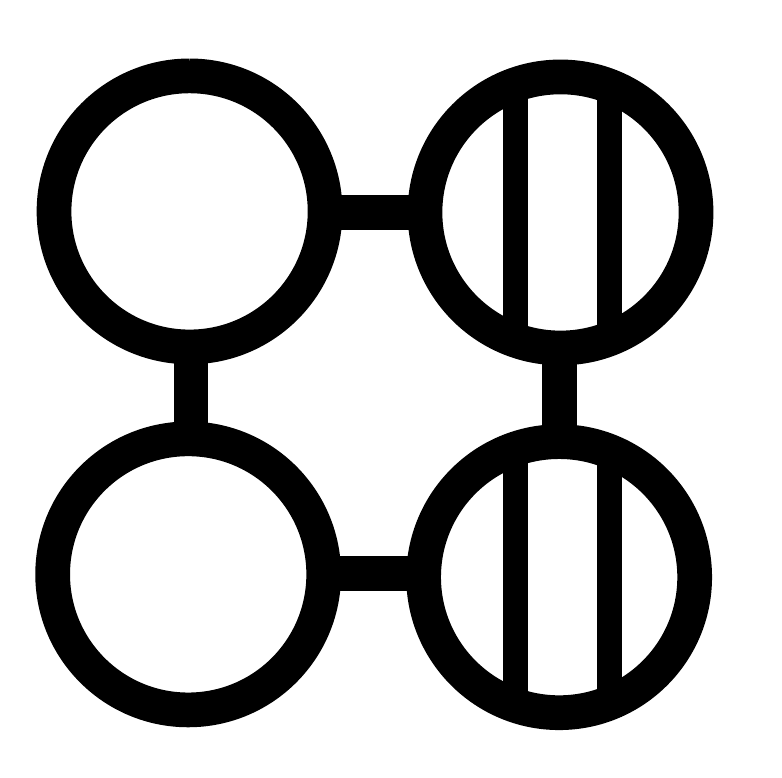}\newline\cartoon{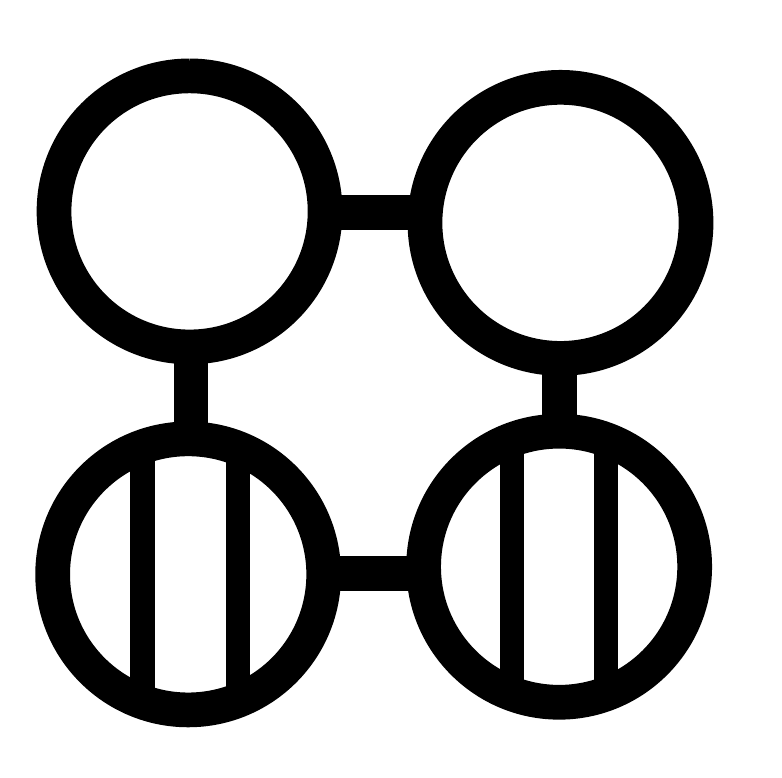} &
    \par$\bar{c}(t)$\par$\bar{c}(t)$&
    \par$\bar{c}(t+f_1T_0)$\par$\bar{c}(t)$&
    \par$\bar{c}(t+f_1T_0)$\par$\bar{c}(t+f_1T_0)$ &  
    \par$\bar{c}(t)$\par$\bar{c}(t+f_1T_0)$\\
\hline
    ${\scriptstyle (D^s_1,1)}_A$\newline${\scriptstyle (D^s_1,1)}_B$& 
    \cartoon{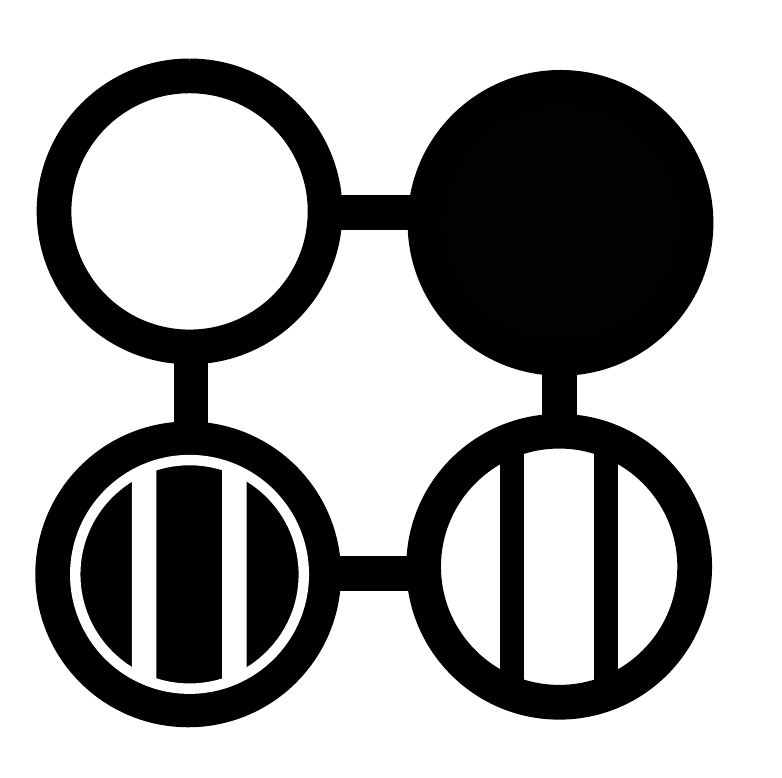}\newline \cartoon{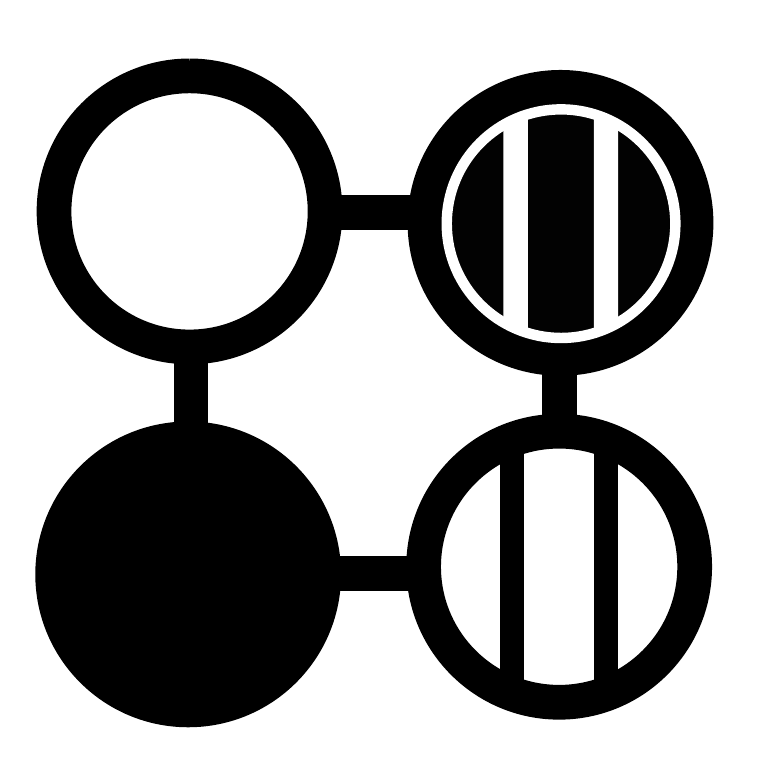}&
    \par$\bar{c}(t)$\par$\bar{c}(t)$&
    \par$\bar{c}(t+\frac{T_0}{2})$\par$\bar{c}(t+(f_1+\frac{1}{2})T_0)$ &
    \par$\bar{c}(t+f_1T_0)$\par$\bar{c}(t+f_1T_0)$ &  
    \par$\bar{c}(t+(f_1+\frac{1}{2})T_0)$\par$\bar{c}(t+\frac{1}{2}T_0)$ \\
\hline
   ${\scriptstyle (Z_2,1)}$ & 
   \cartoon{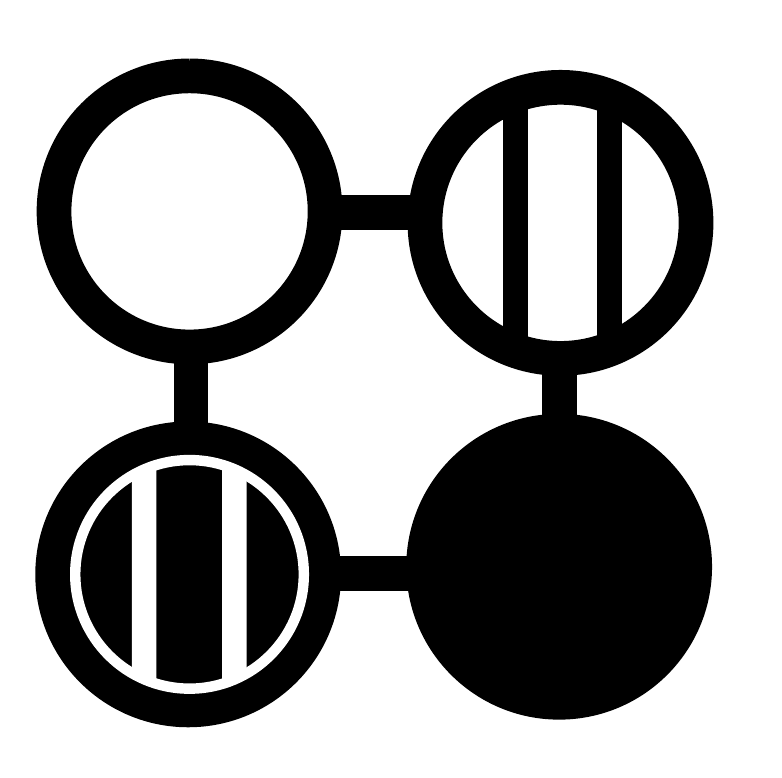} &
   $\bar{c}(t)$ &$\bar{c}(t+f_1T_0)$ & $\bar{c}(t+\frac{T_0}{2})$ &  $\bar{c}(t+(f_1+\frac{1}{2})T_0)$  \\
  \hline
\multicolumn{6}{c}{\text{\small Planar invariant manifolds: }} \\
 \hline
    ${\scriptstyle (D^p_1,D^p_1)}_A$\newline ${\scriptstyle (D^p_1,D^p_1)}_B$ &
    \par\cartoon{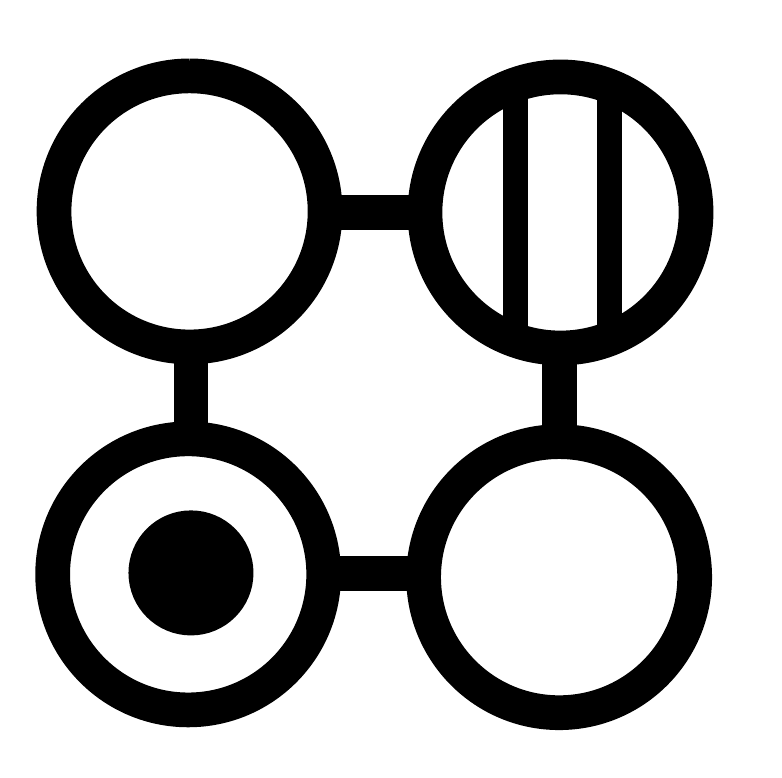}\par\cartoon{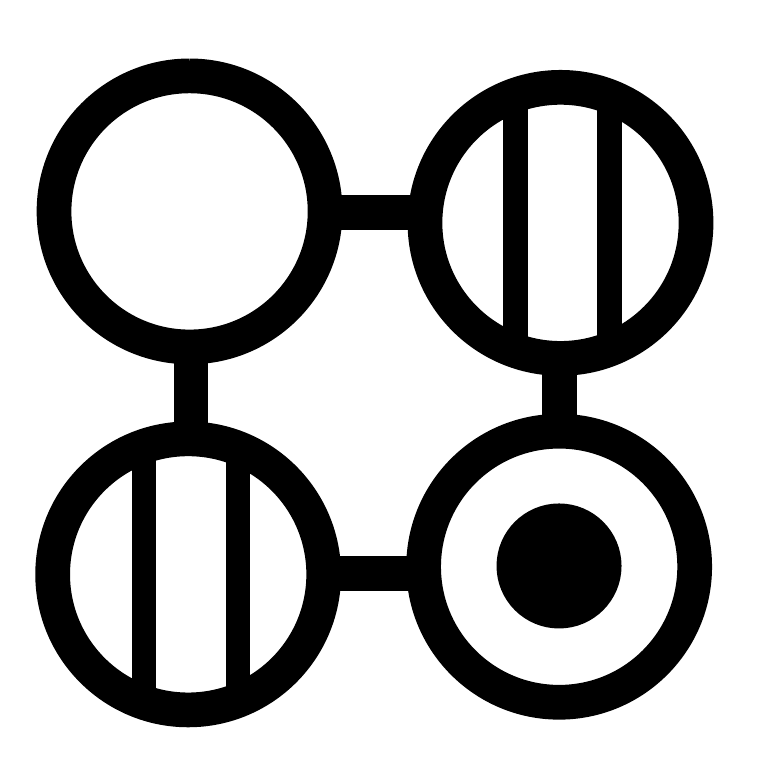}&
    \par$\bar{c}(t)$\par$\bar{c}(t)$&
    \par$\bar{c}(t+f_1 T_0)$\par$\bar{c}(t+f_1 T_0)$ &
    \par$\bar{c}(t)$\par$\bar{c}(t+f_2 T_0)$  &
    \par$\bar{c}(t+f_2 T_0)$\par$\bar{c}(t+f_1 T_0)$  \\
 \hline
 \hline
\end{tabular}
\label{table:lock} 
\end{table}
\end{center}

\par Under the assumption that all reactors are oscillating on the same limit cycle, we can parameterize the time dependent concentrations through the \emph{phase} of that cycle, as proposed by Winfree~\cite{Winfree1967} and widely used in various applications~\cite{Ermentrout2009,Kuramoto1984,Schwemmer2012,Wilson2019a,Norton2019,Monga2019}. In this abstraction, the phase variable naturally progresses linearly from $0$ to $2\pi$ at a frequency $\omega_0=2\pi/T_0$.  Perturbing the chemical concentration of an individual reactor will lead to a modification of the phase that depends on the chemical species that is perturbed and phase of the reactor; this function is called the phase response curve (PRC). The impact on the phase of one reactor due to diffusive coupling from a neighbor  can be summarized through the interaction function, $H$, derived from  convolving the PRC with the diffusive coupling between reactors and averaging over a period~\cite{Kuramoto1984} [\apn{C}]. Notably, this reduction of the chemical model of Eq. \ref{eqn:RDeqhom} to a phase model introduces no new parameters. Best fits between experiment and model are obtained with the interaction function $H$ that arises from a combination of  \ce{Br_2} and \ce{HBrO_2}, as shown in \fig \ref{fig:fitting}(a). This leads to the phase equation: 
\par 
\begin{equation} \frac{d}{dt}\phi_i = \omega_0 + k \sum_{j=1}^4 A_{ij} H(\phi_j - \phi_i)
\label{eqn:phasemodelhom}
\end{equation}
where $k$ the diffusive coupling rate.

\par Noting that the right hand side of Eq.\ref{eqn:phasemodelhom} depends only on phase difference $\theta_{ij}\equiv \phi_i-\phi_j$; we therefore arbitrarily choose the three phase differences $\bar{\theta} = (\theta_{21},\theta_{32},\theta_{43})$ as the new system variables and recast the dynamics accordingly,
\begin{equation}
\frac{d}{dt} \bar{\theta} = \bar{\Psi}(\bar{\theta})
\label{eqn:phasediffmodel}
\end{equation}
with $\bar{\Psi}(\bar{\theta})$ following directly from Eq. \ref{eqn:phasemodelhom}. As each of the phase differences is periodic on $(0,2\pi]$, the state-space is a 3-torus. Although the 3-torus cannot be drawn in three dimensions, it is equivalent to a Cartesian cube with periodic boundaries, allowing visualization of the full dynamics.

\subsection*{Dynamics in the State-space of Phase Differences}
\begin{figure*}
\includegraphics[scale=.1, angle=0]{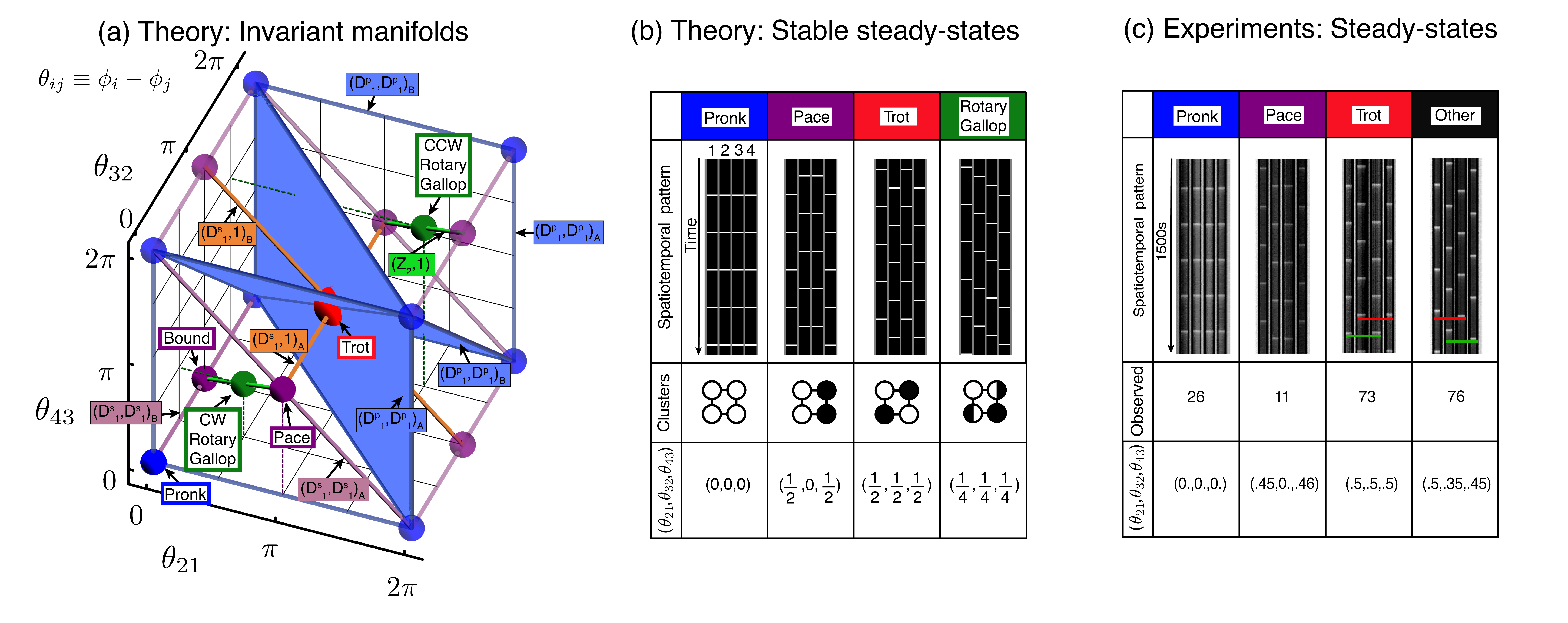}
\caption{ \textbf{(a)} 
The 8 categories of invariant manifolds for a network of 4 nodes with square symmetry predicted by the H/K theory are presented in the state-space of phase differences. There are 4 categories of phase-locked periodic states (points), 3 categories of lines and 1 category of planes.  The state-space is periodic and invariant manifolds that are identical mod $2\pi$ are rendered in translucent colors. \textbf{(b)} (First row) The 4 point manifolds predicted by theory, Pronk, Pace, Trot and Gallop, are represented as space-time plots, (second row) as networks with fixed phase differences and (third row) as a triplet of phase differences, $(\theta_{21},\theta_{32},\theta_{43})$, with units of fraction of a period.  \textbf{(c)} (First row) In experiment, 3 of the 4 phase-locked gaits are observed, while Rotary Gallop is never observed. However, a myriad of states with an unclear correspondence to theory are observed, labeled ``Other''. The ``Other'' states are more similar to Trot than Pronk, Pace or Gallop in terms of space-time plots and phase differences. However, Trot and ``Other'' are qualitatively distinct states because in Trot diagonal nodes oscillate in-phase while in ``Other'' they oscillate with large phase shifts between them, shown by the red and green bars over the space-time plots. (Second row) The number of times a state is observed is recorded and (Third row) the measured phase difference is listed. Videos of experiments shown in Movies S1-S4. 
}
\label{fig:libstates}
\end{figure*}

This new coordinate system transforms the invariant manifolds identified by the H/K theorem in Table \ref{table:lock} to simple, geometric objects: points, lines, and planes, enumerated in Table \ref{table:phase}. This transformation enables the consequences of the H/K theorem on the dynamics in state-space to be visualized in a way that would be impossible in the full chemical model, Eq. \ref{eqn:RDeqhom}. The point invariant manifolds Pronk, Bound, Trot, Rotary Gallop become steady-states in this new frame, rather than high-dimensional limit cycles. We are able to readily classify them as either attractors, repellers or saddles, according to whether the velocity vectors surrounding the steady-state point inward, outward or change sign depending on orientation, respectively \cite{Strogatz1994}.  Beyond the steady-states in the form of points, \fig \ref{fig:libstates}A shows a state-space structured by additional invariant manifolds in the form of lines and planes, all required by the spatiotemporal symmetries of the oscillator network.

\par
\begin{figure*}
\includegraphics[scale=.1, angle=0]{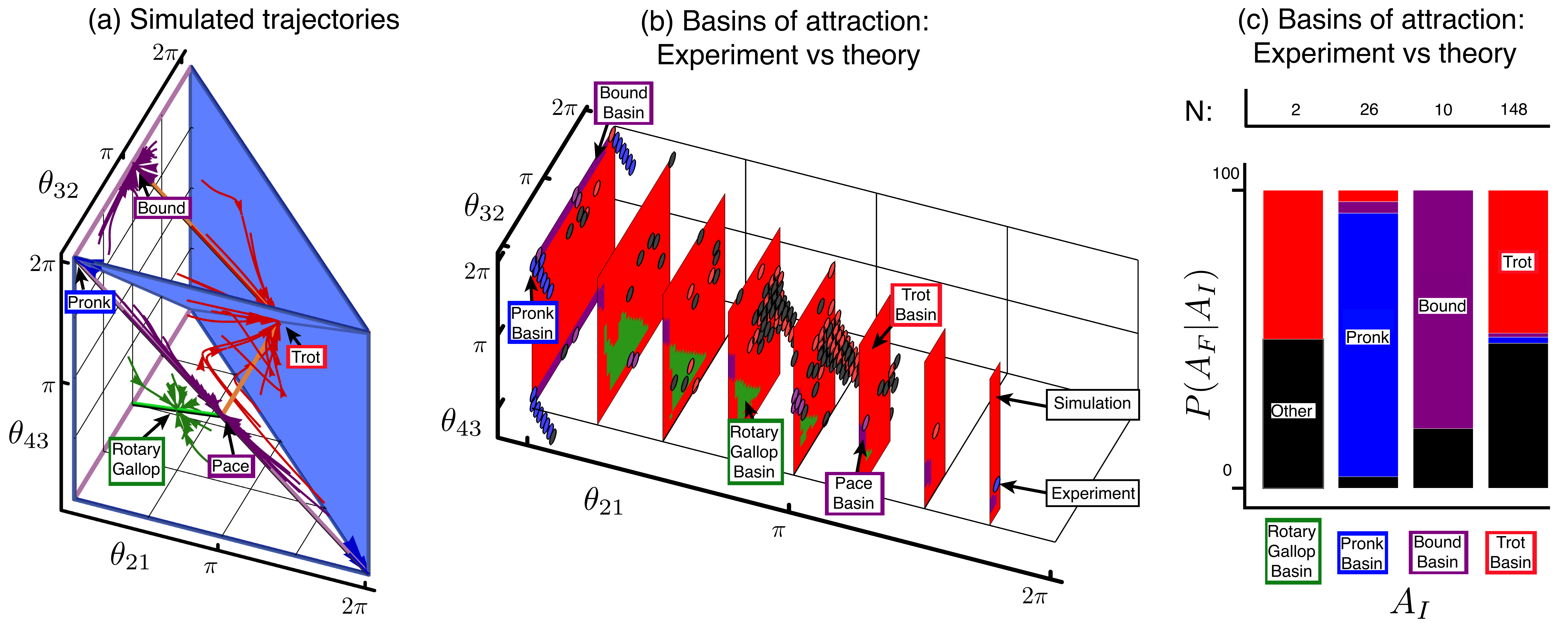}
\caption{Basins of attraction. States are labeled as in Fig. \ref{fig:libstates}.
  \textbf{(a)}  Simulations of eqn \ref{eqn:phasediffmodel} show that all trajectories converge to the H/K point invariant manifolds, Pronk, Bound, Rotary Gallop or Trot, depending on initial condition.  Video of  different perspectives in 3D are shown in movie S5. The corresponding plot of all experiment is shown in movie S6.
  \textbf{(b)}  Basin of attraction for experiments and theory. The theoretical basins of attraction are colormaps computed from $5399$ simulations, including those in (a). For example, if an initial state is colored red, then it will flow to the Trot attractor. The experimental basins of attraction are partially reconstructed by disks located at their initial condition and colored by the attractor to which they converge, with the color code indicated in (c). We artificially shift slightly the experimental points with nearby initial conditions so that they are revealed. Video of different perspectives in 3D are shown in movie S7. 
  \textbf{(c)}  The probability, $P(A_F |A_I)$ , that an experiment which started at a set of phase differences in a given theoretical basin, $A_I$, converges to each of the 4 experimental attractors, $A_F$. The number of observations is listed in the row labeled ``N''. For example, of the 2 initial conditions corresponding to Rotary Gallop, one ended up in the Trot basin and one in the Other basin, but neither went to the predicted basin. In contrast, of the 148 states initially in Trot, 48.7\% went to Other, 2.0\% to Pronk,  1.3\% to Bound and 48.0\% to Trot.  
}
\label{fig:steady}
\end{figure*}

 The dynamical system of Equation [\ref{eqn:phasediffmodel}] predicts that the network is multistable, with the point H/K manifolds forming  competing attractors. We simulated exhaustively the model and observed that each initial condition flows to one of the four categories of phase-locked attractor states; (1) Pronk, (2) Pace/Bound, (3) Trot, and (4) CW/CCW  Gallop [\fig \ref{fig:steady}(a)(b)]. Furthermore, we show in the \ES \sect IIB that the system predicts the six phase-locked states are linearly stable, and thus attractors.

\par
We found that the theoretical model also possesses many unstable, saddle phase-locked steady-states  in addition to the six H/K derived attractors.  In fact, topology predicts the existence of unstable steady-states. The topological index of both attractors and saddles with one attracting direction is +1, while the index of both repellors and saddles with two attracting directions is -1. Topology requires that the sum of the topological indices of all the steady-states must equal 0, the Euler characteristic of a 3-torus, as shown in the \ES \sect IV. Given the six H/K point invariant manifolds are attractors, we conclude there must be at least six unstable steady-states located in the 3-torus to satisfy the required charge neutrality. Moreover, unstable states organize the separatrices between invariant manifolds containing more than one attractor. We numerically searched for the required unstable states and found 158 saddle and 4 unstable steady-states dispersed throughout the state-space shown in  \ES \fig 7.  A notable and unexplained fact is that of the 168 numerically identified phase-locked steady-states, the sole attractors are the six invariant point manifolds required by the H/K theorem.

\begin{center}
\begin{table}[h!]
\centering
\caption{Symmetry required invariant manifolds parameterized by relative phase. $f_1,f_2$, which vary from 0 to $2\pi$. All representations, modulo $2\pi$, are shown.   }
\setlength{\tabcolsep}{5pt} 
\renewcommand{\arraystretch}{1} 
\begin{tabular}{  K{10mm} K{6mm} K{50mm}} 
\hline
\hline
\multicolumn{3}{c}{\text{\small Point invariant manifolds:} } \\
\hline
\textbf{\tiny{Name}} & \textbf{\tiny{Phases}}  & \textbf{\tiny{Hyperplane}} \\ 
\hline
    \text{\tiny{Pronk}} &
    \cartoon{figs/pronk}&
    $(\theta_{21}=\theta_{32}=\theta_{43}=0)$ \\ 
\hline
    \text{\tiny{Trot}}& 
    \cartoon{figs/trot}  &
    $(\theta_{21}=\theta_{32}=\theta_{43}=\pi)$ \\ 
\hline
    \par \text{\tiny Pace}\par \text{\tiny Bound}& 
    \cartoon{figs/bound}\newline\cartoon{figs/pace}&  \par$(\theta_{21}=-\theta_{43}=\pi , \theta_{32} = 0)$ \par$(\theta_{21}=-\theta_{43}=0 , \theta_{32} = \pi)$ \\ 
\hline
    \par \text{\tiny{CW Gallop}}\par \text{\tiny{CCW Gallop}} &
    \cartoon{figs/rgallop}\newline \cartoon{figs/lgallop}& \par$(\theta_{21}=\theta_{32}=\theta_{43}= +\frac{\pi}{2})$ \par$(\theta_{21}=\theta_{32}=\theta_{43}= -\frac{\pi}{2})$ \\ 

\hline
\multicolumn{3}{c}{\text{\small Linear invariant manifolds: }} \\
\hline
    ${\scriptstyle (D^s_1,D^s_1)_A}$\newline${\scriptstyle (D^s_1,D^s_1)_B}$& \cartoon{figs/adjIM}\newline\cartoon{figs/adjIMB} & \par$(\theta_{21}=-\theta_{43}=f_1 , \theta_{32} = 0)$ \par$(\theta_{21}=\theta_{43}=0 , \theta_{32} = f_1)$\\
\hline
   ${\scriptstyle (D^s_1,1)}_A$\newline${\scriptstyle (D^s_1,1)}_B$& \cartoon{figs/adjMIM.pdf}\newline\cartoon{figs/adjMIMB}& 
   \par$(\theta_{21}=\pi , \theta_{32} =f_1 , \theta_{43}=-\theta_{32})$
   \par$(\theta_{21}=f_1 , \theta_{32}=\pi , \theta_{43} = \pi)$ \\
\hline
   ${\scriptstyle (Z_2,1)}$ &
   \cartoon{figs/diagMIM} &
   $(\theta_{21}=f_1 , \theta_{32} = \pi-f_1 , \theta_{43}=f_1 )$\\
\hline
\multicolumn{3}{c}{\text{\small Planar invariant manifolds: }} \\
\hline
    ${\scriptstyle (D^p_1,D^p_1)}_A$\newline${\scriptstyle (D^p_1,D^p_1)}_B$&
    \cartoon{figs/diagIM}\newline \cartoon{figs/diagIMB} & 
    \par$(\theta_{21}=-\theta_{32} , \theta_{32}=f_1 , \theta_{43}=f_2)$
    \par$(\theta_{21}=f_1 , \theta_{32}=f_2 , \theta_{43}=-\theta_{32})$ \\
\hline
\hline
\end{tabular}

\label{table:phase} 
\end{table}
\end{center}

\par We numerically determined the basins of attraction of each attractor by dividing the 3-torus into a fine grid and identifying each initial point with the attractor to which it flowed, as shown in  Fig. \ref{fig:steady}A-B \cite{Pusuluri2020}. The Pronk, Bound and Rotary Gallop basins are smooth, closed volumes while the Trot basin fills the rest of the state-space [\fig \ref{fig:steady}(b)]. The state with largest basin of attraction is Trot, followed by Rotary Gallop, Bound and Pronk. The attraction basin of the Bound  state is anisotropic and aligned with the $(D^s_1,D^s_1)$ invariant manifolds [Fig. \ref{fig:steady}(a),\ref{fig:transient}(a)]. \fig \ref{fig:steady}(a) and \ref{fig:transient}(b) reveal that trajectories remain near the $(D^p_1,D^p_1)$ invariant manifolds as they flow to Trot. Theory predicts that the network's trajectories flowing towards its attractors are constrained and shaped by H/K linear and planar invariant manifolds. 

\par To further elucidate how symmetric invariant manifolds guide and structure dynamics, we focus on the transverse dynamics, namely flows perpendicular to the invariant manifolds. We combined the unique system dynamics, Eqns. \ref{eqn:phasemodelhom} and  \ref{eqn:phasediffmodel}, and the universal structure of the line and plane invariant manifolds to semi-analytically compute the manifolds' transverse Lyapunov exponents, which measure the local attraction or repulsion rate of trajectories to or from them \cite{Ashwin1996,Pecora1998,Pecora2014a}, in terms of the interaction function, $H$ [\apn{E}].  The majority of the domains of the  $(D^s_1,D^s_1)$ and $ (D^p_1,D^p_1)$ manifolds are covered with negative exponents [\fig \ref{fig:transient}(c)]. This causes  trajectories to collapse and remain on these invariant manifolds [\fig \ref{fig:transient}(a)(b)] \cite{Pecora2014a}. While the regions with negative exponents organize the flows along the invariant manifolds, the small regions of positive exponents contain the separatrices of attractors on the same invariant manifolds, shown as saddles in the \ES  \fig 7. In this way, the theory combines the restrictions of symmetry and the unique system dynamics to predict both the basins of attraction of the attractors and the symmetric, clustered transient transitions along the linear and planar invariant manifolds that connect the attractors.

\par The higher order H/K invariant manifolds are not the only invariant manifolds controlling transient flows. Additional invariant manifolds arise in the vicinity of steady-states because linear stability analysis permits an eigenvector decomposition of the dynamics. We see such an invariant manifold exists about the Rotary Gallop attractor as all the surrounding trajectories coalesce into a plane [\fig \ref{fig:steady}(a)]. This invariant manifold is not aligned with any H/K invariant manifold and arises from the unique system dynamics rather than the H/K theorem or symmetry.

\subsection*{Experimental Observations of Dynamics}
\par To compare the geometric framework provided by the theory of equivariant dynamics with experiment, we defined a reactor's phase in reference to the moments of maximum oxidation, corresponding to a maximum in transmitted light intensity [\fig \ref{fig:introdiagram}(d)]. These moments were attributed phase $2\pi$, and phase was defined as the fraction of time spent between two peaks. This allows us to directly compare theory with experiment by measuring the experimental phase differences between 3 sequential, adjacent pairs of reactors $( \theta_{21}, \theta_{32}, \theta_{43} )$, and to characterize the dynamics relative to invariant manifolds and their stability derived from the model. 

\par We conducted 318 experiments of which 186, or 58\%, phase-locked as defined in \apn{B}. During the interval of time soon after the first oscillation and before phase-locking, the oscillation periods of all of the reactors remained similar,  $\pm10\%$. Because the periods of the oscillators are similar, we make the assumption that these reactors are on the same limit cycle.  Experiments were stopped either when the system phase-locked or when 70 oscillations occurred, after which the amount of reactants consumed led to the oscillation periods becoming highly variable. We monitored the initial condition, as well as the full transient trajectory for each of the 186 trajectories on their path to phase locking. This allowed the assessment of whether the trajectories  were constrained by the invariant manifolds and were affected by the transversal stability, as predicted by  theory. 

To classify experimental phase-locked states we measure the distance between the observed state and each of the theoretical attractors using a metric appropriate for a 3-torus [\apn{D}]. Each phase-locked state is classified as the nearest attractor, or as \textit{Other} if they are more than 1.0[rad] from each of the H/K attractors. Of the observed steady-states, a majority of $59\%$ correspond to the predicted Pronk, Bound and Trot point invariant manifolds [\fig \ref{fig:libstates}(c)].  We were initially baffled by the remaining  $41\%$ of the observed phase-locked states as they are located a distance from each of the 4 classes of attractors that exceeds the aforementioned threshold radius [\fig \ref{fig:libstates}(c)], raising the question of their origin as they were not predicted by theory.

\begin{figure*}
\includegraphics[scale=.1, angle=0]{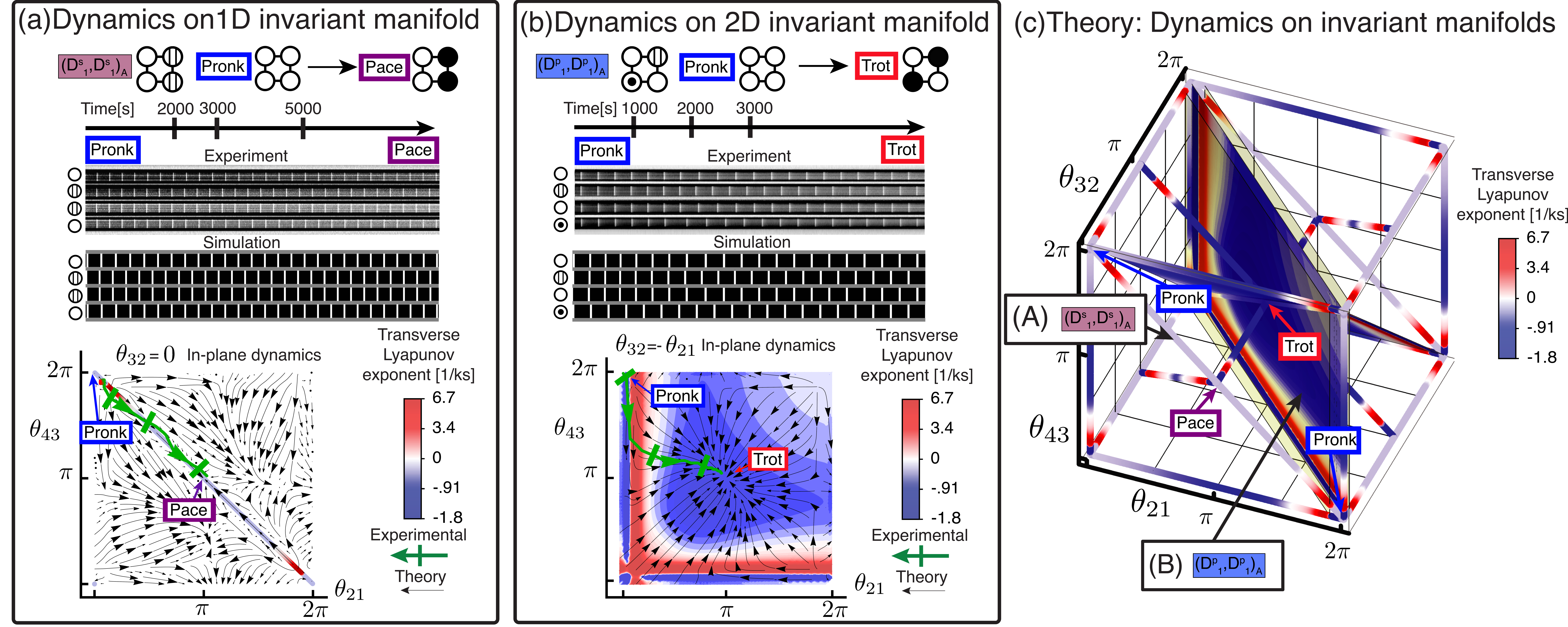}
\caption{ Transient dynamics along higher order H/K invariant manifolds in experiment and simulation.
\textbf{(a)}  Space-time plots from an experiment and simulation with a near-Pronk initial state transitioning to Pace. States form 2 symmetric clusters corresponding to the $ (D^s_1,D^s_1)$ invariant manifold. In the lower panel the experimental trajectory is shown as an arrow  traveling through a 2D slice of state-space superimposed over the theoretical velocity field. Video of experiment synchronized to progression along space-time plot and trajectory in state-space shown in movie S8.
\textbf{(b)} Space-time plots from an experiment and simulation with a near-Pronk initial state transitioning to Trot. The transition corresponds to the $(D^p_1,D^p_1)$ invariant manifold. Video of experiment synchronized to progression along space-time plot and trajectory in state-space shown in movie S9.
\textbf{(c)}
The invariant manifold surfaces attract or repulse in a state-dependent manner. The analytically computed transverse Lyapunov exponents [\apn{E}] are shown via heatmaps. When positive it indicates nearby trajectories are repulsed from the invariant manifold. When negative it indicates attraction. Both the 2D $(D^p_1,D^p_1)$  and 1D $(D^s_1,D^s_1)$ invariant manifolds are largely attracting. Video of 3D perspective of plot in movie S10. 
}
\label{fig:transient}
\end{figure*}

\par Strikingly, \fig \ref{fig:transient}(a)(b) shows that experimental trajectories starting near invariant manifolds $ (D^s_1,D^s_1)$ and $ (D^p_1,D^p_1)$ closely follow the dynamics predicted by the theory. Such a consistency between experimental observations and clustered transient states predicted on the basis of network symmetry alone is quite remarkable. This suggests that the universal properties dictated solely by the symmetry of the system not only predict stationary behaviors of an experimental system, but also constrain transient dynamics from an initial condition to a stationary state.

\par
\begin{figure}
\includegraphics[scale=.1, angle=0]{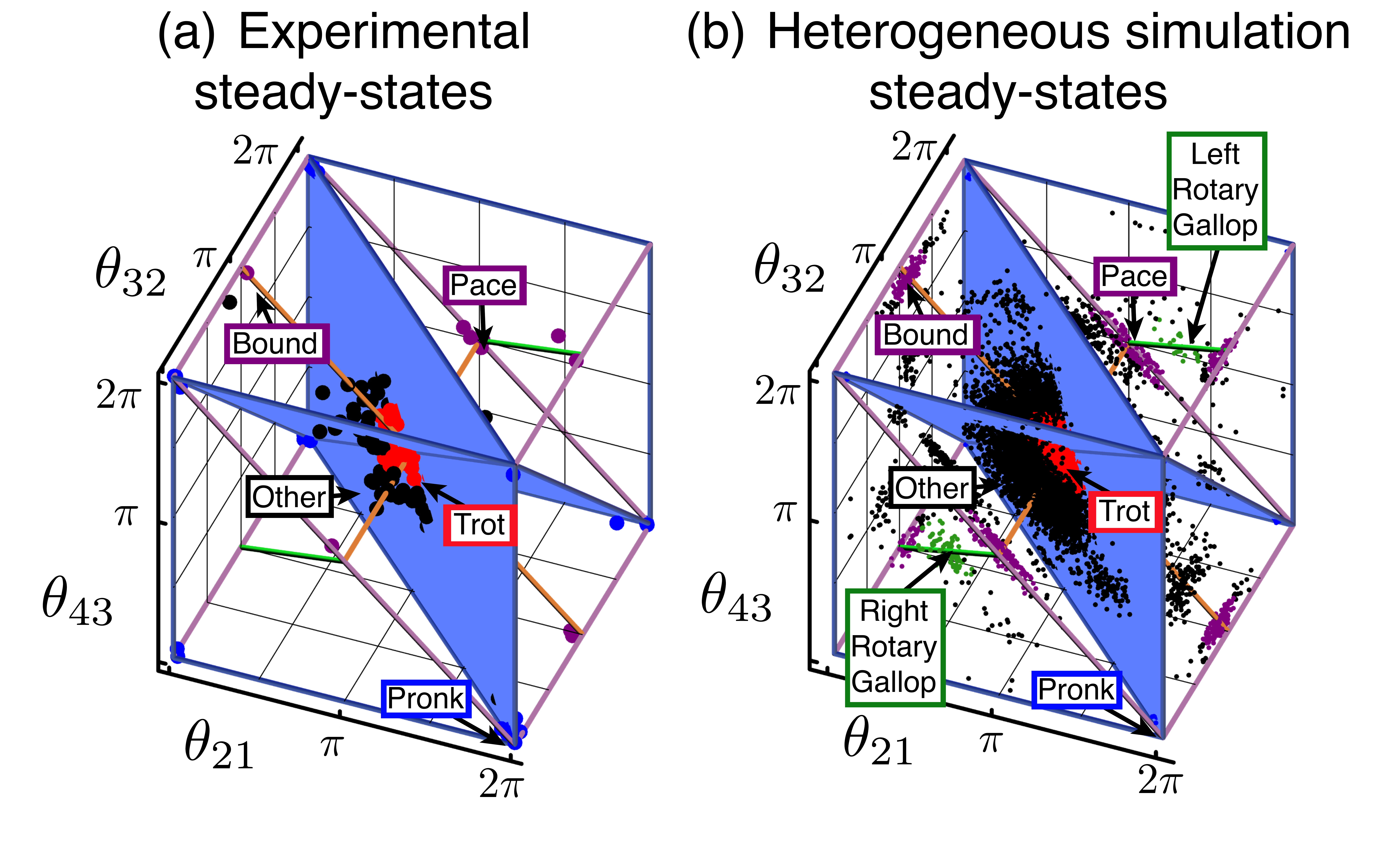}
\caption{ Comparison between experiment and simulation with heterogeneous oscillator frequencies. Phase-locked states are colored by which H/K point manifold they are closest to, or as ``Other'' if they are more than 1.0 [rad] away from each gait. States are labeled as in Fig. \ref{fig:libstates}.
\textbf{(a)} Experiments.
\textbf{(b)} Simulations with heterogeneity. A 3\% percent coefficient of variation in unperturbed frequencies, $\omega_i$, caused a 23-fold reduction in observed Rotary Gallop steady-states compared to the homogeneous simulations in \fig \ref{fig:steady}(a)(b). Videos of 3D perspectives of plots shown in movies S11 and S12, respectively. }
\label{fig:clustering}
\end{figure}

 There are three other noteworthy comparisons to make between theory and experiment. Firstly, we find experiments and theory have similar shaped Pronk, Bound and Trot basins of attraction, as illustrated in \fig \ref{fig:steady}(b)(c). Secondly,  the theoretically predicted Trot attractor is symmetrically surrounded by an extended cloud of phase-locked states, denoted Other, which are significantly far from  Trot, yet within the predicted Trot basin of attraction. These  observations suggest that the experimental Other states are associated with the predicted Trot state, as shown in \fig \ref{fig:steady}(b)(c) and \fig \ref{fig:clustering}(a). Thirdly, the Rotary Gallop states were absent in all our experiments. This is particularly surprising  because the theory attributes to that state a large basin of attraction.

 \subsection*{Inclusion of Slight Heterogeneity in Theory}
 \par A hypothesis to account for these discrepancies between theory and experiment is to consider heterogeneities. Indeed, chemical systems differ from theoretical models in that they are bound to be heterogeneous, and in particular imperfectly symmetrical \cite{Kuramoto1984,Strogatz2001}. For instance, modeling BZ micro-oscillators with small degrees of heterogeneity in reactor chemistry (in turn associated with heterogeneous frequencies), or in reactor volume, led in various situations to better match experimental results \cite{tompkins_testing_2014,Norton2019,Li2015}. We thus tested whether small degrees of heterogeneity between reactors could indeed explain part of the discrepancy between theory and experiments. To this purpose, we  defined a variation of our network model including heterogeneities between reactors, which, using phase reduction [see detail in \apn{C}] led us to analyze a phase model with heterogeneous frequency of type\cite{Schwemmer2012}:
 
\begin{equation} \frac{d }{dt}\phi_i = \omega_i + k \sum_{j=1}^4 A_{ij} H(\phi_j - \phi_i)
\label{eqn:phasemodelhet}
\end{equation}

\par To fit equation \ref{eqn:phasemodelhet} to a trajectory of phase differences measured during an experiment requires a specific set of non-zero unperturbed frequency differences $\omega_i-\omega_j$ [\apn{C}]. Although the best fit unperturbed frequencies are different for each experiment, their statistical distribution fits a Laplacian probability  distribution function corresponding to unperturbed frequencies $\omega_i$ having a percent coefficient of variation of $\pm 3 \%$[\ES \sect II]. Multiple simulations from each initial condition are run using a new sampling from the best fit unperturbed frequency differences probability density function. The steady-state phase differences for phase-locking trajectories observed experimentally and in simulations with heterogeneity are shown in \fig \ref{fig:clustering}.  

\par With the introduction of heterogeneity in the model using probability distributions fitted to the experiments, we observed that the Rotary Gallop state  disappears, just as in experiment [\fig \ref{fig:clustering}]. The percent of phase-locked states which were Rotary Gallop was high in symmetric simulations in which all oscillators had the same frequency, $6.0 \% (N=5,399)$ compared to $0.26 \% (N=34,713)$ in heterogeneous simulations and $0 \% (N=186)$ in experiments. This 23-fold decrease in percentage of Rotary Gallop steady-states is significantly larger than the 1.5, 2.3 and 3.2-fold decreases for Trot, Bound and Pronk. Further, the Trot, Bound, and Pronk states in simulation are clustered in a manner corresponding to experiment [\fig \ref{fig:clustering}(b)]. In particular, we observe  states, termed Other, that form a large cluster of phase-locked states centered about Trot [\fig \ref{fig:clustering}]. We therefore conclude that small levels of heterogeneity can indeed account for the main discrepancies between the theory and the models.

\section*{Discussion}

 Symmetry principles have been used to design network topologies of electronic oscillators that generate desired dynamics \cite{Matheny2019,In2003}. The correspondence between theory and experiment was ascribed to the precision of the fabrication process of these electromechanical devices, which possessed intrinsic frequencies with only 0.001\% percent variation, ensuring the equivalence of each network node and connection \cite{Matheny2019}. While these experiments demonstrated the relevance of equivariant dynamics to experiments, it is unresolved whether the predictions of H/K theorem persist in the biological milieu which never achieve such interchangeability \cite{Winfree1967}.

\par 
Approaching biological systems directly in a similar manner is difficult because the intrinsic dynamics are both complex and unknown. One study found that oscillatory slime mold confined to networks with specified symmetries support aspects of the H/K theory \cite{Takamatsu2001}. The biological oscillators in this system possessed intrinsic frequencies with 10\% percent variation \cite{Takamatsu2000}. Despite being  4 orders of magnitude more heterogeneous than the electronic system, symmetric dynamics emerged. Still, while these results are intriguing, because the underlying chemical dynamics of slime mold are not fully understood and the total number of oscillations are small, it is impossible to assess the stability of observed spatiotemporal patterns, transient dynamics, or the impact of heterogeneity. Thus, minimally complex reaction-diffusion based systems provide an essential linkage between idealized theory and biology. 

\par Our experiments showed both consistencies and discrepancies with the symmetry-based theory. In particular, most phase-locked states were recovered, and tracking trajectories from an initial condition to a phase-locked state showed that the theory not only predicted the steady-states, but also their basins of attraction and, more surprisingly, transient dynamics and transverse stability along invariant manifolds.

However, through combining theory and experiment on a model system,  our results suggest that heterogeneity eliminates some states, but not others, raising the question of understanding which states are the most sensitive to heterogeneity. A natural hypothesis would be that the states disappearing upon addition of heterogeneity would correspond to states which in the absence of heterogeneity, would have weaker stability. Established theory concerning heterogeneity in oscillator networks  predicts the impact of heterogeneity is inversely proportional to the system's linear stability \cite{Norton2019,Skardal2014}. Following this approach, we computed the maximum Lyapunov exponent for each of the phase-locked states. We found that Pronk and Bound have the same maximum Lyapunov exponents, \num{-2 e-4}  $[ \mathrm{s^{-1}} ]$, while Rotary Gallop has a threefold larger value, \num{-6 e-4} $[ \mathrm{s^{-1}}]$ [\ES \fig 6]. In other words, in the absence of heterogeneity, the Rotary Gallop state has a higher magnitude vector field pointing in towards it than the other states. Therefore, interestingly, Rotary Gallop has a stronger stability and a larger basin of attraction relative to Pronk and Bound, and yet appears more sensitive to heterogeneity. This raises the question of characterizing how symmetric states are affected by perturbations, a largely open theoretical question with important applications.

\section*{Conclusion}

\par Understanding how network structure controls spatiotemporal pattern formation remains a central problem in network science. Analysis of spatial network symmetries has led to great progress by illuminating mechanisms behind the emergence of clustered, dynamical states. Specifically, tools for identifying  group orbits ~\cite{Pecora2014a} and equitable partitions ~\cite{Belykh2011, Siddique2018} have been particularly fruitful in systematizing the identification of topology-required clustered states. Here, the low-dimensionality of the representation of a 4-ring network by phase differences allows us to concretely illustrate complex but universal features of the dynamical landscape underpinning the emergence of clusters.

\par We showed that a longstanding theoretical conjecture, that symmetry can dictate function in biological systems, can be used to rationally engineer a spontaneously organizing reaction-diffusion network of Belousov-Zhabotinsky oscillators. Thus, our results offer promise for applying symmetry principles as a tool for designing out-of-equilibrium materials and understanding biological dynamics. Contemporary theory \cite{Golubitsky1999, GolubitskyFull2000} proves that oscillator networks with the same symmetry as the 4-ring we studied are required to share a universal list of phase-locked states and transient dynamics.

By mapping both the high-dimensional chemical model and experimental observations to a 3D state-space of phase differences, we showed that the symmetry required invariant manifolds take on simple forms. Point manifolds are phase-locked states with spatiotemporal patterns that we recognize as quadruped gaits. The higher dimensional manifolds consist of lines and planes in state-space that guide the transient dynamics from one point manifold to another. Additionally, network sub-clusters are sequentially synchronized when convergence to a final spatiotemporal pattern occurs along these manifolds. Thus the H/K theorem imposes a great deal of structure on the phase-locked and transient dynamics of the system that is dependent only on the network's topology and independent of any of the specifics of the oscillators and their coupling. These results therefore provide strong support to the hypothesis that symmetries in chemical or biological neural networks organize functional patterns, such as locomotion \cite{GolubitskyFull2000, Stewart2015}.

\par An important aspect of this study is its exhaustiveness: analyzing a small network allowed a complete application of the H/K theorem. In our integrated reaction-diffusion system, the adjustable parameters were few, and the number of oscillations per trial and the number of trials were large, thereby facilitating a detailed comparison between theory and experiment. Having shown the successes and limitations of the theorem using the general methods of phase-reduction, it provides a framework for analyzing other networks. Most readily comparable will be other networks with polygonal geometry. For larger-scale networks,  it remains in principle possible to numerically employ a similar methodology to predict spatiotemporal patterns by applying the H/K framework in complex networks, for example by using computer-assisted calculations~\cite{Pecora2014a}. However, for large numbers of nodes and symmetries, it may become more practical and meaningful to approximate the network by a continuum and use continuous symmetry groups (e.g., dense lattices could be approximated by planes, or polygons with a large number of nodes by circles, with continuous rotations or translations as symmetries) \cite{Vivancos1995,Bosking1997,Bressloff2001,Bressloff2001a}. Moreover, extensions of the theory will allow prediction of further invariant manifolds, which do not arise directly from the symmetry group of the network \cite{Golubitsky2006,schaub2016, Salova2020}.

 \par  These experiments  raise for the first time the deep theoretical question of how, in spite of this general consistency between symmetry based theories and experiment, that even small levels of heterogeneity have the potential of crucially modifying the dynamics. Two examples are that heterogeneity renders some symmetry-derived states no longer observable and the surprising phenomenon whereby the sensitivity of states to heterogeneous perturbations does not correlate with the strength and size of the basin of attraction of the state in the ideally symmetric system. These results emphasize the importance of assessing the robustness of symmetry-predicted results in the face of heterogeneity. This assessment is essential to validate the application of the symmetry-based theory to biological systems, as well as to guide the design of chemical reaction-diffusion networks to be used in engineered applications, such as soft robotics. 
 
 To date, we know of no theoretical framework addressing the structural stability of H/K’s predictions to heterogeneity that can explain our experimental and  numerical observations. Thus, this work encourages more theoretical and experimental studies such as systematically introducing symmetry breaking by controlling  the degree to which  nodes and connections are distinct, so as to finely characterize the origin of heterogeneity-induced destabilization or vanishing of steady-states\cite{Norton2019}. Our experimental system, used here for the first time to test symmetry based theories in reaction-diffusion networks, is ideally suited for such studies\cite{Litschel2018}. Our results partially reveal  the  complex role of network structure on dynamics, but to articulate fully the engineering principles of network dynamics it remains to elucidate how  heterogeneity impacts  performance. We hypothesize that similarly to the phase-locked and transient dynamics studied here, the impact of network heterogeneity is partially symmetry generic and partially model specific.

\section*{Acknowledgements}

\par We acknowledge financial
support from NSF DMREF-1534890, the U. S. Army Research
Laboratory and the U. S. Army Research Office under contract/
grant number W911NF-16-1-0094, the microfluidics facility of
the NSF MRSEC DMR-2011486, and the Swartz Foundation Grants 2017-6 and 2018-6.

\par IH performed all experiments, data analysis, and simulations. MN guided all of the work. Experimental design by IH, MN, and SF. BC, CS, MN, JT, and IH contributed to understanding theoretical role of symmetry in system. MM helped fabricate microfluidic chips. IH, MN, JT, and SF wrote the manuscript.

\par Simulations were performed using Brandeis University's High Performance Computing Cluster which is partially funded by DMR-MRSEC 2011486

We acknowledge R\'{e}mi Boros, Youssef Fahmy, and Amanda Chisholm for their preliminary experiments on 4 ring networks. We are grateful for Jan Engelbrecht and Rennie Mirollo for their spirited discussions on dynamical systems theory.

\section*{Appendix A: Experimental Methods}
\label{sec:AA}
\subsection{Network Fabrication}
\par The microfluidic reaction-diffusion network was made out of four adjacent reactors embedded in polydimethylsiloxane (PDMS). The reactors are formed out of divots in PDMS, forming effective buckets, which can be filled and then sealed all together by a piece of glass, forming a common lid. We manufactured these divots using a soft lithographic process in which PDMS is cured while pressed against an inverse (positive) of the divots made out of a photoresist deposited onto a silicon wafer. This was performed as previously published \cite{Litschel2018}, with the exception of one adaptation described below. This generates a glass microscope slide coated with many reactors organized into networks of four reactors, shown in \fig \ref{fig:introdiagram}(b)(c) and \ES \fig 1 and 2(a).

\par The dimensions chosen for the network allow for robust coupling of four nodes in ring topology. By adjusting the sizes and distances between reactors we found that rectangular reactor dimensions 62\si{\micro \metre} x 62\si{\micro \metre} x 30\si{\micro \metre} (L x W x H)  with side-to-side distance 26 \si{\micro \metre} resulted in strong coupling. The network reactors are organized in a 2 by 2 grid [\fig \ref{fig:introdiagram}(b)(c)] in such a way that nearest neighbor reactors possess much more shared surface area relative next-nearest neighbors across the diagonal. This results in a ring-like connectivity, where coupling between nearest neighbors is stronger than across the diagonal. The rectangle of BZ surrounding the network [\fig \ref{fig:introdiagram} (c)] is forced into a steady-state, setting the concentration of chemicals surrounding the network. During each experiment we observe nine or sixteen strongly coupled, individual networks, separated from one another by controlled barriers  [\ES \fig 1(b)(c)]. 

The only alteration of the procedure in fabricating the PDMS networks published \cite{Litschel2018} was to change the way in which the PDMS was pressed and cured – instead of a 15\si{\kg} lead brick applied for 12 hours followed by baking in a 70C oven, we used a thermal press applying 90-113\si{\kg} set at 70C for 2.5 hours. This was found to: a) reproducibly keep the size of the layer of PDMS underneath sample less than 2 \si{\micro \metre} [\ES \fig 1(d)] and b) decrease the probability that the silicon wafer breaks per use. 

\subsection{Sample Holders}
\par In a previous work the PDMS reactors had BZ sealed inside of them and were loaded into a microscope using an acrylic plastic clamp \cite{Litschel2018}. This clamp did not control the temperature of the BZ. However, the frequency of BZ oscillations depends on temperature \cite{Bansagi2009}.
\par To maximize experimental reproducibility, we created a clamp that controlled sample temperatures to within 0.1\si{\celsius}. The clamp's temperature is controlled through a thermistor that measures the temperature of the clamp nearby the sample [\ES \fig 2], 2 Peltier (TEC) devices [\ES \fig 2], and PID feedback between them mediated by an Arduino. The sample is robustly driven to the clamp's temperature because the clamp possesses a large thermal mass relative the sample and large thermal contact area with the sample [\ES \fig 2]. During all trials samples were kept at $22.0$ \si{\celsius}.

\par We seal samples in the temperature-controlled clamp exactly the same way as with the previous, plastic clamp \cite{Litschel2018}, described in \apn{A} Protocol. 
\subsection{BZ Chemical Preparation}
\par 
The BZ loaded into the microfluidic network is first mixed outside the microfluidic device. A .24\si{\milli \liter} volume of photo-sensitive BZ is prepared by sequentially adding equal $60$\si{\micro \liter} volumes of  Sulfuric acid, Sodium Bromide, Malonic acid, Sodium Bromate, Ferroin then Tris(2,2'-bipyridyl)dichlororuthenium(II)hexahydrate to an Eppendorf tube, then mixing it with a Vortex mixer. Note that during the sequential pipetting of the chemicals, upon adding the Sodium Bromate, the solution converts from colorless to a vivid, transparent yellow for 15 seconds before returning to a colorless state. The volumes output by the pipette used had a measured percent coefficient of variance of $1.2 \% $. The concentrations of the reagents in the final .24\si{\milli \liter} mixture, and ultimately in the individual BZ microreactors, are in Table \ref{table:chems}.

\begin{table}\centering
\caption{Final experimental chemical conditions in reactors:}
\setlength{\tabcolsep}{0pt} 
\renewcommand{\arraystretch}{1} 
\begin{tabular}{K{3.5cm} K{3.3cm} K{2cm}}
 \hline
 \hline
  \text{Chemical}  &\text{Molecular Formula} & \text{Concentration} \si{ \milli \Molar} \\
  \hline
 Sulfuric Acid & \ce{H_2SO_4}& $80$\\ 
 \hline
  Sodium Bromide &\ce{NaBr} & $25$  \\ 
  \hline
  Malonic Acid & \ce{C_3H_4O_4}& $400$ \\ 
 \hline
  Sodium Bromate &\ce{NaBrO_3} & $288$  \\ 
 \hline
 Ferroin & \ce{C_{36}H_{24}FeN_6O_4S} & $3$\\ 
 \hline
  Tris(2,2’-bipyridyl) dichlororuthenium(II) hexahydrate & \ce{C_{30}H_{24}Cl_2N_6Ru. 6H_2O} & $1.2$\\ 
 \hline
\end{tabular}
\label{table:chems}
\end{table}

\subsection{Optics}

\par We measured the chemical state of the reactors through measuring their absorbance of green light. Ferroin's absorbance of green light changes drastically between its oxidized and reduced state. The green light is filtered to  $515\pm10$\si{\nano \metre}, to avoid exciting  the photocatalyst, \ce{Ru(bipy)_3}.

\par Light perturbation, used to set boundary conditions and initial conditions were set by projection of patterned blue light onto the sample, selectively exciting the photocatalyst. As in previous works \cite{Litschel2018,tompkins_testing_2014,Delgado2011}, the patterned blue light was periodically turned off and on at a high frequency to allow accurate measurement of absorbance of Ferroin, with period $2$ seconds and duty cycle of 50$\%$. During some of the experiments the blue light was homogenized by directly replacing the sample with a CCD and using feedback between projected signals and the measured values to minimize measured variability, as previous published \cite{Sheehy2020}.
\par Boundary conditions were applied by shining light on the rectangle surrounding the network at an intensity that completely inhibit oscillations in it. 
\par Initial conditions were set by applying light to the reactors by inhibiting all reactors with light for 300-600 seconds. Then, the light was turned off at different times from each of the reactors, thus causing them to resume oscillating at different times. The success rate of hitting target initial conditions far from Trot or Pronk was low.

\par The light intensity of sample illumination was measured by placing a power meter in the sample plane, the results are similar to in previous \cite{Delgado2011} work: Intensity of blue  light applied to boundaries: $0.3\pm.04$ \si{\milli \watt \per \centi \metre \squared}, Intensity of blue  light applied to reactors during initial condition setting: $1\pm0.2 $\si{\milli \watt \per \centi \metre \squared}, Intensity of blue applied light when projector blank/black: $0.09\pm0.009 $\si{\milli \watt \per \centi \metre \squared}, Intensity of 515nm green sample illumination: $\sim$0.1\si{\milli \watt \per \centi \metre \squared}. Errors, in standard deviations, express variance in average illumination across the whole sample field of view across all experiments, not the variance across the field of view in individual experiments

\subsection{Protocol}
\par The protocol for an experiment is as follows:
\begin{enumerate}
    \item PDMS chip, reentrant window, and O-ring [\ES \fig 2(a)] are cleaned with isopropyl alcohol, deionized water, and dried with compressed air. They are left under petri dishes to prevent dust accumulation. 
    \item A small batch of BZ solution is prepared as detailed earlier in \apn{A}. Solution is left in a dark chamber.
    \item The PDMS chip is plasma treated for 3 minutes at 400\si{\milli\bar} in ambient atmosphere. 
    \item The BZ solution is then pipetted into the networks of interest in the PDMS chip as shown in depth in supplementary movie S7 of \cite{Litschel2018}.
    \item Now, with the reentrant window placed approximately above a feature of networks covered by BZ, the reentrant window must be secured more firmly and precisely. While viewing the sample using a stereomicroscope with green filtered transmission illumination, the thumbscrews  [\ES \fig 2] are slowly turned, clamping the device. We alternated tightening them in a zig-zag pattern, with each tightening of a screw being roughly a 1/8 or less rotation. During this process any bubbles which are present in the reactors should decrease in size until they are invisible. Once all reactors are surrounded by dark outlines \ES \fig 1(a), there are no shearing distortions to the network, and there are no bubbles, this process is halted.
    \item The clamp and the network with BZ sealed into it are then left in a dark, room temperature chamber until it has been 40 minute since the BZ was initially mixed in step 2, typically 20 minutes.
    \item The clamp is then loaded into the projection illumination microscope \ES \fig 3(c). Then, a MATLAB code with GUI is used to align a projected pattern onto the sample \ES \fig 3(b) and initiate temperature control. 
    \item Light is projected onto boundaries and sets initial conditions of networks as described earlier in \apn{A}. Data is gathered for between 3000 and 24000\si{\second}, $\sim$ 10 and 81 periods of oscillation of each reactor. 
    \item In a few experiments a second attempt at setting initial conditions was made. 
\end{enumerate}

\section*{Appendix B: Phase-Locked Criteria}
\label{sec:AB}

\par To identify phase-locked states in experiments we require that $\frac{d}{dt}(\phi_i-\phi_j)$ is almost zero and is not accelerating, $\frac{d^2}{dt^2}(\phi_i-\phi_j)$ is also small. 
The algorithm used:
\begin{enumerate}
    \item Calculate the three phase differences versus time: $(\theta_{21},\theta_{32},\theta_{43})$
    \item Lowpass them to form: $(\overline{\theta_{21}},\overline{\theta_{32}},\overline{\theta_{43}})$ 
    \item Find the longest region in the $(\overline{\theta_{21}},\overline{\theta_{32}},\overline{\theta_{43}})$ time series when their velocities at below as threshold: $$|\frac{d}{dt}\overline{\theta_{21}}|,|\frac{d}{dt}\overline{\theta_{32}}|,|\frac{d}{dt}\overline{\theta_{43}}|<2.5\times10^{-4} \left[ \frac{\text{rad}}{\text{s}} \right]$$ and the average acceleration is also below a threshold: $$\frac{d}{dt}\frac{1}{3}(|\frac{d}{dt}\overline{\theta_{21}}|+|\frac{d}{dt}\overline{\theta_{32}}|+|\frac{d}{dt}\overline{\theta_{43}}|)<9\times10^{-8}\left[ \frac{\text{rad}}{\text{s}^2} \right]$$
    \item If the longest region is 5 or more periods of oscillation of all oscillators (1500 seconds), we consider the experiment to be phase-locked. 
\end{enumerate}

\section*{Appendix C: Best Fit Model}
\label{sec:AC}

\subsection{Physical Models of BZ Microoscillators}
The chemical concentration oscillations of an isolated BZ microreactor are accurately modeled by the reaction kinetics derived for macroscopic reactors \cite{vanag_model_2009,Sheehy2020}. Denoting concentrations, $\bar{c}=(x,y,z,u)$ where $x=$[\ce{HBrO_2}], $y=$[\ce{Br-}], $z=$[Oxidized catalyst], $u=$[\ce{Br_2}]:
\begin{equation} 
\frac{d}{dt} \bar{c} = \bar{R}(\bar{c})
\label{eqn:1reactor}
\end{equation}

$$\bar{R}( \begin{bmatrix} x \\ y \\ z \\ u \end{bmatrix}) =\begin{bmatrix} R_x(x,y,z) \\ R_y(x,y,z,u) \\ R_z(x,z) \\ R_u(x,y,u) \end{bmatrix}   $$

\begin{gather}
  R_x(x,y,z) = k_2y-k_1xy-2k_3x^2+\frac{k_4x(c_o-z)}{(c_o-z+c_{min})}  \nonumber \\ R_y(x,y,z) = -2k_2y+k_7u+k_9z-3k_1xy-k_3x^2\nonumber\\ +\frac{k_I(c_o-z)}{(\frac{b_C}{b}+1)} \nonumber \\
   R_z(x,z) = -(k_9+k_{10})z +2\frac{k_4x(c_o-z)}{(c_o-z+c_{min})} +\frac{k_I(c_o-z)}{(\frac{b_C}{b}+1)}  \nonumber \\
   R_u(x,y,u) = k_2y-k_7u+2k_1xy+k_3x^2 \nonumber
\end{gather}

\begin{table}\centering
\caption{Simulation parameters known:}
\begin{tabular}{  K{1cm} K{1cm} K{5cm} K{1cm} } 
 \hline
 \hline
 \multicolumn{4}{c}{\text{Reagent concentrations:} } \\
 \hline
 \text{} & \text{Description } & \text{Value} & \text{Unit} \\
 \hline
 $a$   & Bromate  & $288$ & \si{\milli \Molar} \\ 
 \hline
 $m$    & Malonic acid & $400$ & \si{\milli \Molar} \\ 
 \hline
 $c_o$ & Total metal ion catalyst & $4.2$ & \si{\milli \Molar} \\ 
 \hline
 $h$   & Protons & $160$ & \si{m\Molar} \\ 
 \hline
  $b$    & Bromomalonic acid & $0.12*m$ & \si{\milli \Molar} \\ 
 \hline
 \multicolumn{4}{c}{\text{Reagent rates and relevant constants:} } \\
 \hline
  \text{} & \text{ } & \text{Value} & \text{Unit} \\
 \hline
  $k_1$    & \text{} & \num{2e6}$h$ & \si{\per \Molar \per \second } \\
  \hline
  $k_2$    & \text{} & $2h^2a$ & \si{\per \second } \\ 
    \hline
  $k_3$    &  \text{} & $\num{3e3}$ &  \si{\per \Molar \per \second }\\ 
    \hline
  $k_4$    &  & $42ha$ &\si{\per \second }  \\ 
    \hline
  $k_5$    &  & $\num{5e9}h$ & \si{\per \Molar \per \second }  \\ 
    \hline
  $k_6$    &  & $10$ & \si{\per \second } \\ 
    \hline
  $k_7$    &  & $29m$ &\si{\per \second }  \\
    \hline
  $k_8$    &  & $9.3m$ &\si{\per \second } \\ 
    \hline
  $k_9$    &  & $b$ &\si{\per \second }  \\ 
    \hline
  $k_{10}$    &  & $0.05m$ &\si{\per \second }  \\ 
    \hline
  $k_r$    &  & $\num{2e8}$ &\si{\per \Molar \per \second }\\ 
    \hline
  $k_{red}$    &  & $\num{5e6}$ & \si{\per \Molar \per \second } \\ 
    \hline
  $k_{I}$    &  & $0$ &\si{\per \second } \\ 
    \hline
  $b_C$    &  & $0.05$ & \si{\Molar} \\ 
    \hline
  $c_{min}$    &  & $\sqrt{2 k_r (k_9+k_{10})c_o/k_{red}^2}$ & \si{\Molar} \\ 
    \hline
\end{tabular}
\end{table}

The exchange of chemicals between adjacent BZ microreactors is limited because the PDMS between them is apolar. \ce{Br_2} is the only apolar intermediate of the BZ reaction which is certainly soluble in and diffusing through the PDMS between reactors. \ce{HBrO_2} may also be soluble to a lesser degree \cite{tompkins_testing_2014,li_combined_2014,Torbensen2017a,Torbensen2017a}. Because of the short separations between reactors $\mathcal{O} (10\mu m )$, the  diffusion of these chemicals between reactors is assumed to be quasi-static \cite{Toiya2010,Norton2019}. This results in a simple form of linear difference coupling between a pair of BZ microreactors, where $k$ is the diffusive coupling rate of \ce{Br_2} and $k_e$ is the ratio of the diffusive coupling rate of \ce{HBrO_2} relative \ce{Br_2}: 
\begin{gather}
\frac{d}{dt} \bar{c}_1 = \bar{R}(\bar{c}_1)+\mu(\bar{c}_2-\bar{c}_1)  \nonumber \\
\frac{d}{dt} \bar{c}_2 = \bar{R}(\bar{c}_2)+\mu(\bar{c}_1-\bar{c}_2) \nonumber \\ \mu =
\begin{bmatrix} k_e k & 0 & 0& 0 \\
 0 & 0 & 0& 0 \\
  0 & 0 & 0& 0 \\
   0 & 0 & 0& k 
\end{bmatrix} \nonumber 
\label{eqn:coupling}
\end{gather}

\par To simulate our results, we adapt the above, established model. Firstly, we model the ring-like topology by adding strong diffusive coupling between nearest neighbors and weaker coupling coupling across the diagonal. Secondly, we also permit a small degree of heterogeneity in reaction rates within the reactors, representing: i) differences in reagent concentrations, ii) differences in powers of light hitting photo-sensitive reactors, and iii) slightly different boundary conditions. The model is then a function of: a) $k$, the diffusive coupling rate of \ce{Br_2}, b) $k_e$, the ratio of the coupling rate of \ce{HBrO_2} relative \ce{Br_2} c) $f$, the ratio of diagonal to nearest neighbor coupling, d) differences in local reactor reaction rates:

 \begin{equation}  
\frac{d}{dt }\bar{c}_i= \bar{R}_o(\bar{c}_i)+\epsilon \bar{r}_i(\bar{c}_i) + \sum_{j \neq i}^4 A_{ij}\mu(\bar{c}_j-\bar{c}_i)
\label{eqn:RDeq}
\end{equation}
 
 $$A_{ij} = \begin{bmatrix} 0 & 1 & f & 1 \\
                                     1 & 0 & 1 & f \\
                                     f  & 1 & 0 & 1\\
                                     1 & f  & 0 & 0\\ 
 \end{bmatrix}; \mu =  \begin{bmatrix} k_e k & 0 & 0 & 0 \\
                                     0 & 0 & 0 & 0\\
                                     0 & 0 & 0 & 0\\
                                     0 & 0 & 0 & k\\ 
 \end{bmatrix};$$

\par To ease comparison to experiments and qualitatively understand the role of free parameters, we use established methods \cite{Schwemmer2012,Ermentrout2009,Wilson2019a}, described later in this section, to reduce the model above \eqn \ref{eqn:RDeq} to the phase model \eqn \ref{eqn:phasemodelhet}. The reduced model \eqn \ref{eqn:phasemodelhet} predicts the phase dynamics of the reactors relative one another as a function of the parameters: a) $k$ which scales the interaction function $H$, b)  $k_e$ which determines the shape of the $H$ function \fig \ref{fig:fitting}(a), c)  $f$ which changes the adjacency matrix, d) differences in local reactors which change intrinsic frequencies $\omega_i$.

We then construct a 3D model of the evolution of phase-differences between adjacent nodes in the network. Letting $\theta_{ij} \equiv \phi_i - \phi_j $ and $\Delta \omega_{ij} \equiv \omega_i - \omega_j$
\begin{gather} \frac{d }{dt}\theta_{ij} = \Delta \omega_{ij} +k\biggl[\sum_{k \neq i}^4 A_{ik}H(\phi_k- \phi_i)   \nonumber\\
 -\sum_{k \neq j}^4 A_{jk}H(\phi_k - \phi_j)\biggr] \\
\label{eqn:generalphasediff}
\end{gather}

\begin{equation}\frac{d}{dt}\bar{\theta} = 
\begin{bmatrix} \frac{d}{dt}\theta_{21} \\
                                     \frac{d}{dt}\theta_{32} \\
                                     \frac{d}{dt}\theta_{43} \\
 \end{bmatrix} = \bar{\Psi}(\bar{\theta})
 \label{eqn:phasediffmodelgen}
\end{equation}

\par This gives us a model which is expected to qualitatively capture the dynamics of the system and allow us to better understand it. Since the parameters of the model, $f$, $k$ and so on, cannot be independently measured, the model must be empirically fit to experimental data. However, the best fit model reveals important, nontrivial insights into our particular symmetric 4 node network and BZ microoscillator networks in general. 

\par The $H$ function is shown in \fig \ref{fig:fitting}(a). We ran simulations with the $H$ function saved as a Chebychev function \cite{Driscoll2014} using MATLAB's ODE45 with relative and absolute tolerances of  \num{1e-10}.

\subsection{Fitting to Experiments}

\begin{figure*}
\includegraphics[scale=.1, angle=0]{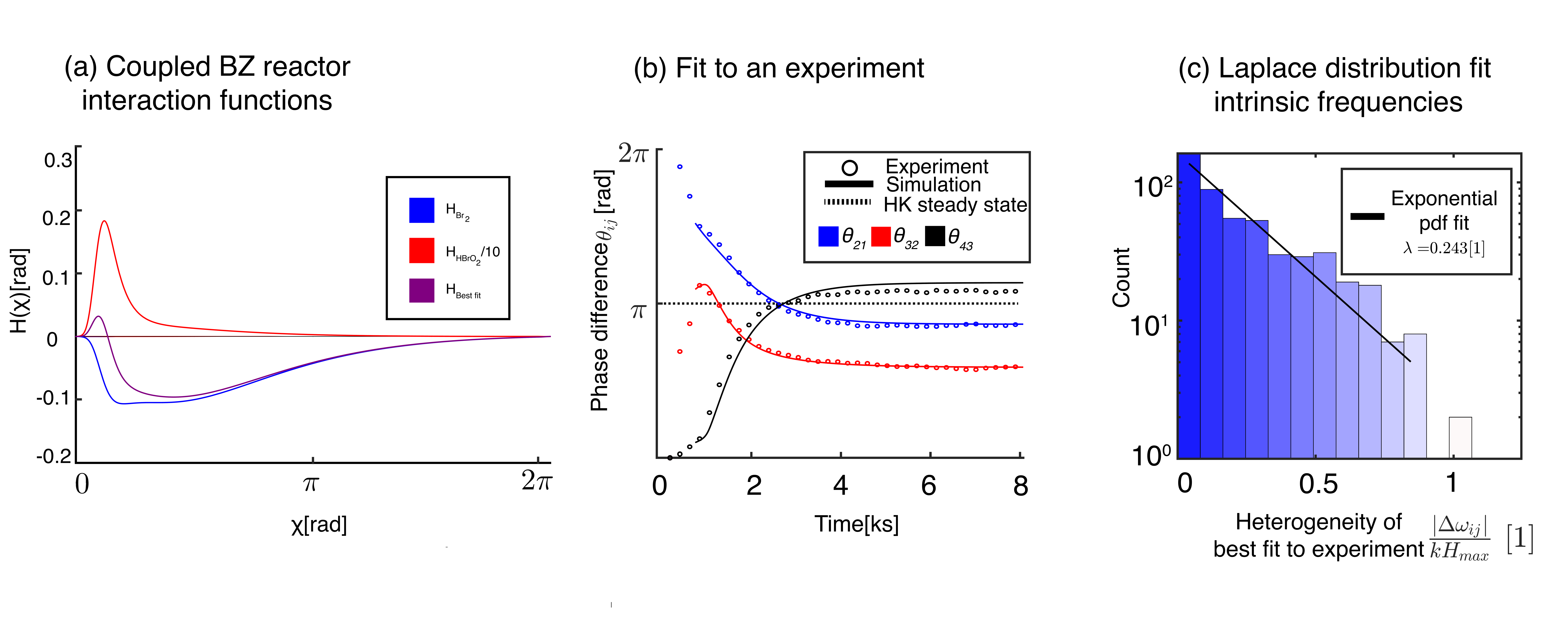}
\caption{
\textbf{(a)} The interaction functions of coupled BZ oscillators. $H_{\text{Br}_2}$ is the interaction functions of purely \ce{Br_2} coupled reactors. $H_{\text{HBrO}_2}$ is the interaction functions of purely \ce{HBrO_2} coupled reactor. In the model of our 4 node network \eqn \ref{eqn:phasediffmodelgen}, when  $k_e$ is not zero the interaction function is $H(\chi)=H_{\text{Br}_2}(\chi)+k_eH_{\text{HBrO}_2}(\chi)$. Since $H_{\text{HBrO}_2}$ is largely 0 except for $\chi$ between $0$ and $\pi/2$, $k_e$ adds a bump of phase advance near $0$, without 
affecting the otherwise phase delaying dynamics due to \ce{Br_2}. $H_{\text{Best fit}}$ corresponds to the interaction function which best fits experiments $k_e=0.05$. 
\textbf{(b)} Example experimental trajectory converging to steady-state (circles), and best fit phase model (solid line). The fit model required slight heterogeneity $\Delta \omega_{ij}\neq0$. The steady-state phase difference $(\pi,\pi,\pi)$ of the phase model, without best fit heterogeneity, at dashed horizontal line. This is the H/K predicted point invariant manifold nearest to the experimental trajectory. 
\textbf{ (c)} Distribution of best fit nondimensionalized intrinsic frequency differences. They are nondimensionalized by dividing them by best fit coupling rate $k$ times the amplitude of the interaction function $H$, $H_{max}$. 
}
\label{fig:fitting}
\end{figure*}

\par The model \eqn \ref{eqn:phasediffmodelgen} was fitted to each experimental time series of phase differences versus time [\fig \ref{fig:fitting}(b)] using nonlinear regression. The nonlinear regression was performed on each individual experiment by constraining a simulation to start from an experimental initial condition, then optimizing the simulation's parameters to reduce the squared error between its trajectory and the experiment's trajectory using matlab 2019b's surrogate optimization. Specifically, the initial condition of phase difference of an experiment was set to be the phase differences when the reactors meet the similar frequency threshold defined in \apn{A}, shown in \ES \fig 4(a)(b). The final point in an experimental trajectory used in a fitting was half way between when the phase-locked condition was met and when it was lost [\ES \fig 4(a)(c)] or the end of the experiment if it did not unlock. 
\par The identified best fit parameters were: $\Delta \omega_{ij}\neq0, k = 1.8\times10^{-2}\pm2.7\times10^{-3}$ [\si{\per \second}], $f=0$[1] and $k_e=.05$[1]. The intrinsic frequency differences $\Delta \omega_{ij}$ obeyed a Laplacian distribution $\rho(\Delta \omega_{ij}) = \frac{1}{2 b } \exp(-\frac{|\Delta \omega_{ij}-\mu|}{b})$ [\fig \ref{fig:fitting}(c)]. The Laplacian distribution of intrinsic frequencies has a mean, $\mu$, of 0 and rate parameter, $b$, of $2\pi7\times 10^{-5}$[\si{\radian \per \second}]. Further, the fitting required some excitatory coupling $k_e=.05[1]$ and no diagonal coupling $f=0[1]$. The best fit values and the values used in simulations shown in Table \ref{table:fitparams}.

\begin{table}\centering
\caption{Simulation parameters fitted: In 'Fit values' and 'Values used in theory' a single number represents the number fit or used in simulations. If in a fit or simulation values were randomly distributed, the form of the distribution is described by 'G' or 'L'.  'G($\mu,\sigma$)' represents a Gaussian probability density function with mean $\mu$ and standard deviation $\sigma$. A 'L($\mu,b$)' represents a Laplacian  probability density function with mean $\mu$ and rate parameter $b$. }
\begin{tabular}{  K{1.2cm} K{3.5cm} K{2.5cm} K{8mm}} 
 \hline
 \hline
Parameter  & Fit values& Values used in theory & Unit \\
  \hline
 \multicolumn{4}{ c}{\text{Coupling:} } \\
 \hline
 $k$    & G$(1.8\times10^{-2},2.7\times10^{-3})$ & $2\times10^{-2}$ &\si{\per \second}\\ 
 \hline
 $k_e$   &  0.05 & 0.05 &1\\ 
 \hline
  $f$   &  0 & 0 &1\\ 
 \hline
  \multicolumn{4}{c}{\text{Chemical heterogeneity:} } \\
 \hline
 $\Delta \omega_{ij}$   & L$(0,2\pi7\times10^{-5})$ &
 L$(0, 2\pi8\times10^{-5})$ &\si{\radian \per \second}\\ 
\hline
\end{tabular}
 \label{table:fitparams}
\end{table}

\par For a discussion of the values of these parameters measured in previous works and their possible physical relevance, please see \ES \sect II.

\subsection{Best Fit Simulations with Heterogeneity}

\par To determine the impact of experimentally realistic heterogeneity on the model, we ran simulations of \eqn \ref{eqn:phasemodelhet} with the  experimental best fit parameters in Table \ref{table:fitparams}. Specifically, heterogeneity in intrinsic frequencies in simulations $\Delta \omega_{ij}$ were drawn from the distribution which fits experiments, a Laplacian distribution defined in Table \ref{table:fitparams}, with mean 0 and rate parameter $2\pi$\num{8e-5}[\si{\radian \per \second}], while all other parameters are the exact, constant value enumerated in the third row of Table \ref{table:fitparams}.

\par In running simulations from a dense set of initial conditions, each initial condition had 7 simulations initialized from it with independent resamplings of frequency heterogeneity. We thus could observe the impact of heterogeneity throughout state-space by sampling the distribution of heterogeneity in all regions of state-space. The result of 34,713 such simulations are shown in \fig \ref{fig:clustering}(b).

\subsection{Computing Phase Model Reduction}
\par We use established methods \cite{Schwemmer2012,Wilson2019a,Ermentrout2009} to reduce the reaction-diffusion model of our 4 reactor network \eqn \ref{eqn:RDeq} to the phase model \eqn \ref{eqn:phasemodelhet}. In this framework, we first determine the phase-dependent phase shift of an uncoupled BZ reactor induced by sudden, small additions of its chemical species. We then compute the fluxes of chemical species between a pair of diffusively coupled reactors as a function of their relative phases. By combining these, we determine the rate of phase shift of a reactor as a function of its phase difference with a neighbor.

\par If chemicals are added to a BZ reactor on its limit cycle, the time until it next spikes will be changed. These time shifts, or phase shifts, depend on amount of chemicals added as well as the time along a period of oscillation, or phase, they are added at. We quantify the phase shifts of an uncoupled reactor to a perturbation at a given phase in terms of phase response curves (PRC). The PRC, $Q$, is a set of curves of phase shift, $\Delta \phi$,  as a function of  a chemical perturbation's chemical concentration, $\Delta \bar{c}$, and  the phase, $\phi$,  at which it is added at. Specifically, it allows calculation of $\Delta \phi = Q(\phi).\Delta \bar{c}$. the case of the 4D model of a BZ reaction \ref{eqn:1reactor}, $Q:\mathbb{R}^4\mapsto \mathbb{S}^1$ is a set of 4 curves with units of phase per chemical concentrations [\si{\radian \per \Molar }]. We compute $Q$ in the limit of infinitesimal perturbations $|\Delta \bar{c}|\to 0$ by the adjoint method \cite{Schwemmer2012,Wilson2019a,Ermentrout2009}. This allows us to interpret $Q$ as the total derivative $Q(\phi) = \nabla_{\bar{c}}\phi$. 

\par Noting that rate of change of phase of a reactor is $\frac{d}{dt}\phi_i$ and is a sum of constant term, its intrinsic frequency $\omega_i$, and a time-varying term due to its coupling with its neighbors depending on their relative phases:  $$\frac{d}{dt}\phi_i=\omega_i + \sum_j A_{ij} F(\phi_i,\phi_j)$$
 Using the PRC we can then compute the rate of change of phase of a reactor $F$ [\si{\radian \per \second}] with respect to an incident, relative phase dependent flux due to coupling $\bar{g}(\phi_i,\phi_j)$ [\si{\Molar \per \second}] using the PRC and the chain rule:
\begin{equation}
F(\phi_i,\phi_j)= Q(\phi)\bar{g}(\phi_i,\phi_j) \approx \nabla_{\bar{c}}\phi \bar{g}(\phi_i,\phi_j)
\end{equation}

Letting the unperturbed limit cycle of a BZ reactor be parametrized $\bar{c}_{LC}(\phi)$ and denoting the neighboring reactor $j$, we see from our diffusive coupling \eqn \ref{eqn:RDeq} that the specific form of phase dependent flux neighbors experience $\bar{g}(\phi_i,\phi_j) = \mu\left[\bar{c}_{LC}(\phi_j)-\bar{c}_{LC}(\phi_i)\right]$ and it follows that:
\begin{equation}
 F(\phi_i,\phi_j)= Q(\phi)\mu\left[\bar{c}_{LC}(\phi_j)-\bar{c}_{LC}(\phi_i)\right]
\end{equation}

Note that $F$ is a linear combination of flux due to \ce{Br_2}, $u$, and \ce{HBrO_2}, $x$: 

\begin{gather}
F(\phi_i,\phi_j)= k [
Q_{u}(u_{LC}(\phi_j)-u_{LC}(\phi_i)) \nonumber \\
+
k_e Q_{x}(x_{LC}(\phi_j)-x_{LC}(\phi_i)) ] \nonumber \\
\end{gather}

Since the change of relative phase differences is small during a period of reactor oscillation, we compute $F(\phi_j,\phi_i)$ averaged over a cycle  \cite{Schwemmer2012,Wilson2019a,Ermentrout2009}. Doing so transforms it into a function of relative phase difference $H(\phi_j-\phi_i)$. Since $F$ is scaled by $k$, we define $H$ such that it must be used by explicit scaling by $k$, the diffusive coupling rate of \ce{Br_2} [\si{s}$^{-1}$] in eqn \ref{eqn:phasemodelhet}:
\begin{equation}
H(\phi_j-\phi_i) \equiv k^{-1}(2\pi)^{-1}\int_0^{2\pi} F(\alpha,
\alpha+\phi_j-\phi_i)d\alpha \end{equation}
We express the interaction function $H$ as the sum of two distinct, separately calculable terms, because $F$ is a linear combination of two functions, one for \ce{Br_2} and one for \ce{HBrO_2}:
\begin{gather}
H(\phi_j-\phi_i) = H_{\ce{Br_2}}(\phi_j-\phi_i)
+k_eH_{\ce{HBrO_2}}(\phi_j-\phi_i) 
\end{gather}
\section*{Appendix D: Metric of Distance}
\label{sec:AD}

\par To measure distances between two points in the state-space of the 3D phase difference dynamics, both experimentally and in simulations, we found a surprising function $d(\bar{\theta},\bar{\theta}')$ was required. For a given pair of points $\bar{\theta}$ and $\bar{\theta}'$ $d$ is calculated by the following algorithm:
\begin{enumerate}
    \item Consider two points in the state-space:
$ \bar{\theta}'=(\theta_{21}',\theta_{32}',\theta_{43}' )   \text{ and }  \bar{\theta}=(\theta_{21},\theta_{32},\theta_{43}) $
    \item Compute phase difference of fourth edge, which is completely determined by the other three
$ \bar{\theta}'_f=(\theta_{21}',\theta_{32}',\theta_{43}' ,\theta_{21}'+\theta_{32}'+\theta_{43}'  )=(\theta_{21}',\theta_{32}',\theta_{43}' ,\theta_{41}')  \text{ and }  \bar{\theta}_f=(\theta_{21},\theta_{32},\theta_{43},\theta_{41})   $
    \item Define a vector of phase difference between states with $\angle$ being complex, or phasor, angle $\theta_{diff_j} = \angle \exp (i*(\theta'_{f_j}-\theta_{f_j}))$
    \item Let the distance between $\bar{\theta}'$ and $\bar{\theta}$ be the Euclidean norm of the 4D phase difference vector $d(\bar{\theta},\bar{\theta}')=|\bar{\theta}_{diff}|_2$
\end{enumerate}

\section*{Appendix E: Computing Transverse Lyapunov Exponents of Invariant Manifolds}
\label{sec:AE}

\begin{figure}
\includegraphics[scale=.15, angle=0]{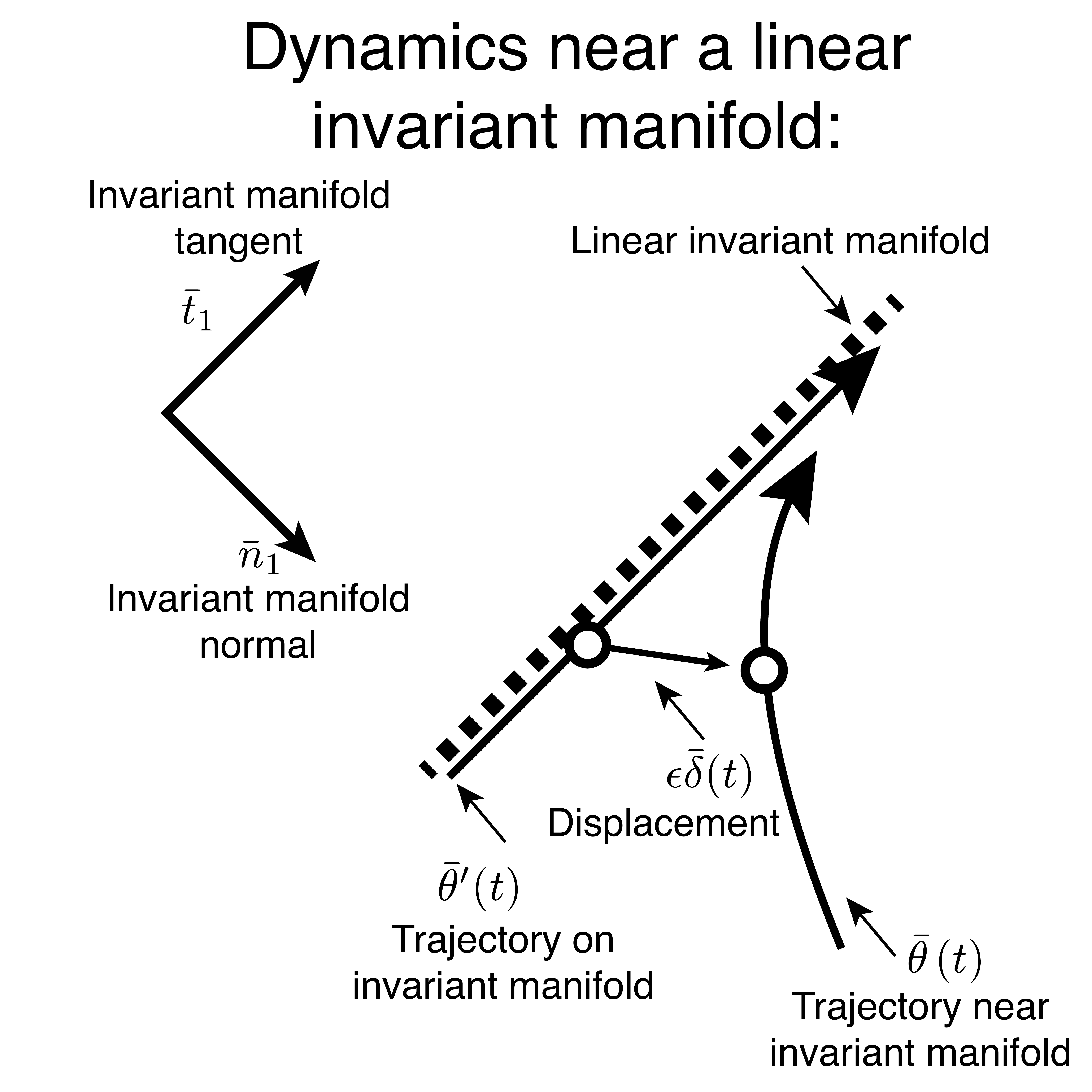}
\caption{
 Diagram of flows nearby a linear invariant manifold. Two example trajectories $\bar{\theta}'$ and $\bar{\theta}$, which are flows in the state-space of \eqn \ref{eqn:phasediffmodelgen}. Note that because $\bar{\theta}'$ is on an invariant manifold, it must continue flowing tangent the invariant manifold indefinitely. However, the trajectory near the invariant manifold $\bar{\theta}$ is not limited in this way. We can consider the displacement between these two trajectories as a function of time $\epsilon\bar{\delta}(t)$. In the case shown, the displacement is shrinking, indicating that the nonlinear flow about the linear invariant manifold near $\bar{\theta}'$ is converging towards it. To quantify this behavior more broadly, we utilize he fact the normal component of the displacement will take on a form $\bar{n}_1^T\epsilon\bar{\delta}(t)\propto \exp(\lambda t)$. The exponential scale $\lambda$, called the transverse Lyapunov exponent, when negative indicates attraction, when positive repulsion.
}
\label{fig:IMex}
\end{figure}
\par \par The maximum transverse Lyapunov exponent (MTLE) of the higher order invariant manifolds describe whether trajectories collapse or diverge from them. If a higher order invariant manifold has a negative MTLE, trajectories will converge towards it \cite{Ashwin1996}. In our particular system, we found the MTLE were often easy to calculate and easy to qualitatively understand. 

\par  The MTLE of the invariant manifolds can be computed at the phase model level using the well known method of linearizing the dynamics \eqn  \ref{eqn:phasediffmodelgen} about the invariant manifolds \cite{Ashwin1996,Pecora1998,Belykh2002,Russo2011,Pecora2014a}. We explicitly list how we performed this procedure in our case in which the invariant manifolds are linear hyperplanes.
 
\par We begin by considering whether a point $\bar{\theta}'$ perturbed off an invariant manifold, displaced by $\epsilon \bar{\delta}$ from point $\bar{\theta}'$, converges back to or diverges from the manifold [\fig \ref{fig:IMex}]. The dynamics of the perturbed point take on a simple form:
\begin{gather}
    \frac{d}{dt} \bar{\theta} = \bar{\Psi}(\bar{\theta}) =\bar{\Psi}(\bar{\theta}'+\epsilon \bar{\delta}) \nonumber \\
    = \bar{\Psi}(\bar{\theta}')+\epsilon \nabla_{\bar{x}} \bar{\Psi}(\bar{x})|_{\bar{x}=\bar{\theta}'}\bar{\delta} + \mathbb{O}(\epsilon^2)
\end{gather}
It is convenient to consider the displacement between the perturbed solution and the solution on the invariant manifold:

\begin{gather}
    \frac{d}{dt} \bar{\delta}=\frac{d}{dt}\bar{\theta}-\frac{d}{dt} \bar{\theta}' = \bar{\Psi}(\bar{\theta}) - \bar{\Psi}(\bar{\theta}')   \nonumber \\
    =\epsilon \nabla_{\bar{x}} \bar{\Psi}(\bar{x})|_{\bar{x}=\bar{\theta}'}\bar{\delta} + \mathbb{O}(\epsilon^2)
\end{gather}
We can determine if the component of the perturbation along a given direction normal to the invariant manifold, $\bar{n}$, grows or decays of the displacement off of the invariant manifold. Letting $J(\bar{x}')\equiv \nabla_{\bar{x}} \bar{\Psi}(\bar{x})|_{\bar{x}=\bar{x}'} $, the rate of change of the normal distance, $\bar{n}^T\bar{\delta}$ is given by:
\begin{gather}
    \frac{d}{dt} (\bar{n}^T \bar{\delta})\approx\bar{n}^T\epsilon J(\bar{\theta}')\bar{\delta} 
    \label{eqn:normdynscomplex}
\end{gather}
\par To precisely determine the exponential timescale with which the normal components of the perturbed trajectory grows or shrinks, we use a unitary transformation matrix $P$ to projects a vector of phase differences from the canonical basis into a basis aligned with a given invariant manifold:  Letting $\bar{t}_i$ be tangent the invariant manifold and $\bar{n}_i$ be perpendicular, $P = \left[\bar{t}_1...\bar{n}_1... \right]^T $ s.t. if $\bar{\theta}=c_{t1}\bar{t}_1+c_{t2}\bar{t}_2...+c_{n_1}\bar{n}_1+..$. then $P.\bar{\theta}  = (c_{t1},c_{t2},...c_{n1}...)$. 

\par Crucially, points on the invariant manifold have no normal components $\forall_j c_{nj}=0$ thus $P.\bar{\theta}'  = (c_{t1},c_{t2},...0,..)$. Similarly, on the invariant manifold the velocity field $P.\bar{\Psi}(\bar{\theta}') = \frac{d}{dt}( c_{t1},...c_{n1},...)$ must have all its normal component be 0, $\forall j \frac{d}{dt}c_{nj}=0$, as by definition trajectories cannot flow out of, or transverse, an invariant manifold \fig \ref{fig:IMex}.

\par We can now consider the growth or decay of perturbations in or out of a manifold. Let $\bar{\xi}=P.\bar{\delta}$:
\begin{equation}
    \frac{d}{dt} \bar{\xi} = \frac{d}{dt}(P.\bar{\delta}) = PJ\bar{\delta}=PJP^{-1}\bar{\xi}
\end{equation}

\par Letting $J'=PJP^{-1}$: $\frac{d}{dt}\bar{\xi} = J'\bar{\xi}$ - a linear time-varying equation. Since the dynamics of the system $\bar{\xi}$ normal to the invariant manifold are all 0, independent on the location along the manifold spanned by the tangent, it follows that the block dynamics in $J'$ corresponding to normal components will be independent of those for tangent components \cite{Ashwin1996}. The maximum real eigenvalue of this block of normal dynamics gives an approximate bound on the exponential growth or decay of the trajectory's distance normal an invariant manifold and is called the maximum transverse Lyapunov exponent.

\subsection{Algorithm}
\par The algorithm, detailed at length above, is as follows:
\begin{enumerate}
  \item Choose an invariant manifold  $M$, of dimension $k$, in an $n$ node network.
  \item Determine $k$ orthonormal vectors which span $M$ and label them the tangent vectors $T$.  Determine a set $N$ of $n-1-k$ vectors orthonormal one another and $T$. We use the Gramm-Schmitt procedure.  
  \item Compute a unitary transformation matrix $P$, with the first columns composed of invariant manifold tangents, then followed by normals. 
  \item Compute the Jacobian of your nonlinear flow at points on the invariant manifold $J(\bar{\theta}')$, written explicitly as a function of location on the invariant manifold $\bar{\theta}'$.
  \item Transform the Jacobian into its tangent and normal components using $P$ via $P.J(\bar{\theta}').P^{-1}=J'(\bar{\theta}')$.
  \item Extract the block of $J'$ that contains the decoupled transverse dynamics - the columns and rows corresponding to normal components. 
  \item Compute the maximum real eigenvalue of the normal block $\lambda(\bar{\theta}')$, which is the \textbf{max Transverse Lyapunov Exponent}
\end{enumerate}

\par An example of executing this algorithm for the $(D_1^p,D_1^p)$ invariant manifold of the 4 ring is in \ES \sect IIIA. Generic expressions of block of $J'$ for all invariant manifolds of a broad class of 4 ring networks are computed in terms of first derivatives of $H$ in \ES Table I. A comparison of the MTLE of a ring of 4 Kuramoto oscillators to our system is presented in \ES \fig 8.

\bibliography{FourRingPaperAV1}

\end{document}



\title{The Symmetry Basis of Pattern Formation in Reaction-Diffusion Networks  \\
\textit{Electronic Supplement}}

\author{\text{Ian Hunter}}
\thanks{These two authors contributed equally}
\affiliation{
Brandeis University Physics, Waltham MA, 02453 USA
}%

\author{Michael M. Norton}
\thanks{These two authors contributed equally}
\affiliation{%
Center for Neural Engineering, Department of Engineering Science and Mechanics, The Pennsylvania State University, University Park, PA 16802 USA
}%

\author{Bolun Chen}
\affiliation{
Brandeis University Volen National Center for
Complex Systems, Waltham MA, 02453 USA}
\affiliation{
 Department of Physics Boston University, Boston MA, 02215 USA
}%
\author{\text{Chris Simonetti}}
\affiliation{
Brandeis University Physics, Waltham MA, 02453 USA
}
\author{\text{Maria Eleni Moustaka}}
\affiliation{
Brandeis University Physics, Waltham MA, 02453 USA
}%
\author{Jonathan Touboul}
\affiliation{
Brandeis University Volen National Center for
Complex Systems, Waltham MA, 02453 USA}
\affiliation{
Brandeis University Mathematics Department}

\author{\text{ Seth Fraden}}
\email[To whom correspondence should be addressed:]{fraden@brandeis.edu}
\affiliation{
Brandeis University Physics, Waltham MA, 02453 USA 
}%

\date{\today}

\keywords{Nonlinear Dynamics,Nonlinear Dynamics,Complex Systems
}

\maketitle



\maketitle

\section{Movies}
\begin{itemize}
\item{Movie S1: Video of an experimental network converging to the Trot attractor: Initially three reactors flash at a similar times, with the fourth flashing out of phase relative to them. As time goes on the system converges to a nearly ideal realization of the Trot state. The raw video of the experiment is shown alongside a space-time plot of the video and the calculated phase differences between reactors 1 and 2, 2 and 3, and 3 and 4. The phase differences are denoted $\theta_{ij}\equiv \phi_i - \phi_j$.
Link to video: \url{https://tinyurl.com/3te6ex37}
}
\item{Movie S2: Video of an experimental network in the Pace attractor: The experiment starts in a Pace pattern and proceeds to stay in it during the rest of the experiment. The raw video of the experiment is shown alongside a space-time plot of the video and the calculated phase differences between reactors 1 and 2, 2 and 3, and 3 and 4. 
Link to video: \url{https://tinyurl.com/2d2p4d9u}
}
\item{Movie S3: Video of an experimental network in the Pronk attractor: The experiment starts in a Pronk state and proceeds to stay in it the during rest of the experiment. The raw video of the experiment is shown alongside a space-time plot of the video and the calculated phase differences between reactors 1 and 2, 2 and 3, and 3 and 4. 
Link to video: \url{https://tinyurl.com/apku3xcx}
}
\item{Movie S4: Video of an experimental network in the Other attractor: The experiment starts and stays in a state in which flashing is initiated at the top left, and proceeds to the bottom right reactor in a Z-pattern. Since this does not resemble any H/K state, it is categorized as Other. The raw video of the experiment is shown alongside a space-time plot of the video and the calculated phase differences between reactors 1 and 2, 2 and 3, and 3 and 4.  
Link to video: \url{https://tinyurl.com/dupdjf9k}
}
\item{Movie S5: Video of rotation about 3D plot in manuscript \fig 3(a) which shows the state-space of the theoretical model. The video starts from the perspective of the \fig 3A, then rotates about it to help readers understand the 3D structure of the state-space. 
Link to video: \url{https://tinyurl.com/63zsevcu}
}
\item{Movie S6: Video of rotation about a 3D plot of the same state-space as \fig 3(a) except with experimental data. Two panels are shown: the left being the trajectories that converged Trot, Bound, and Pronk and the right being the trajectories that converged Other.  The left panel, has clear similarity to the simulated state-space in Movie S5.
Link to video: \url{https://tinyurl.com/c7n7tkmj}
}
\item{Movie S7: Video of rotation about 3D plot in manuscript \fig 3(b) which shows basins of attraction of the theoretical model and  experiments. The video starts from the perspective of the \fig 3(b), then rotates about it to help readers understand the 3D structure of the basins of attraction.
Link to video: \url{https://tinyurl.com/ye6thfme}
}
\item{Movie S8: Annotated video of experiment converging from Pronk to Pace \fig 4(a). The raw video of the experiment is shown alongside a space-time plot of the video and evolution of the phase differences between reactors moving through the state-space of the network. As during this experiment the 2nd and 3rd reactor oscillate in-phase, only the 2D slice of the state-space in which the trajectory is confined to, $\theta_{32}=0$, is shown. Within this view of state-space the 1D H/K predicted invariant manifold $(D^s_1,D^s_1)$ is shown and is colored by its maximum transverse lyapunov exponent, as in the manuscript \fig 4(a)(c). The experimental trajectory noisily moves about Pronk before transitioning to Pace directly along the invariant manifold.
Link to video: \url{https://tinyurl.com/tsyw6xk3}
}
\item{Movie S9: Annotated video of experiment converging from Pronk to Trot \fig 4(b).  The raw video of the experiment is shown alongside a space-time plot of the video and evolution of the phase differences between reactors moving through the state-space of the network. As during this experiment the 1st and 3rd reactor oscillate in-phase, only the 2D slice of the state-space in which the trajectory is confined to, $\theta_{32}=-\theta_{21}$ where $\phi_1=\phi_3$, is shown. Within this view of state-space the 2D H/K predicted invariant manifold $(D^p_1,D^p_1)$ is shown and is colored by its maximum transverse lyapunov exponent, as in the manuscript \fig 4(b)(c). The experimental trajectory starts near Pronk before efficiently transitioning to Trot, directly following the in-plane velocity field of the 2D the invariant manifold.
Link to video: \url{https://tinyurl.com/354k63ma}
}
\item{Movie S10: Video of rotation about 3D plot in manuscript \fig 4(c) which shows theoretical stability of invariant manifolds.  The video starts from the perspective of the \fig 4(c), then rotates about it to help readers understand the structure of the invariant manifold stability.
Link to video: \url{https://tinyurl.com/5yh6m6k5}
}
\item{Movie S11: Video of rotation about 3D plot in manuscript \fig 5(a) which shows the steady-states of experiments. The video starts from the perspective of the \fig 5(a), then rotates about it to help readers understand the 3D structure of the steady-states in the state-space of the experiments. It can be seen clearly that the Other states are distributed anisotropically about Trot, such that they are more widely spread across a plane spanned by the two linear, orange invariant manifolds, $(D^s_1,1)$. 
Link to video: \url{https://tinyurl.com/2ykm5796}
}
\item{Movie S12: Video of rotation about 3D plot in manuscript \fig 5(b) which shows the steady-states of heterogeneous simulations. The video starts from the perspective of the \fig 5(b), then rotates about it to help readers understand the 3D structure of the steady-states in the state-space of the heterogeneous model. The Other states are distributed anisotropically about Trot, such that they are more widely spread across a plane spanned by the two linear, orange invariant manifolds, $(D^s_1,1)$.
Link to video: \url{https://tinyurl.com/3mvttf4d}
}
\end{itemize}

\begin{figure*}[h!]
\includegraphics[scale=.1, angle=0]{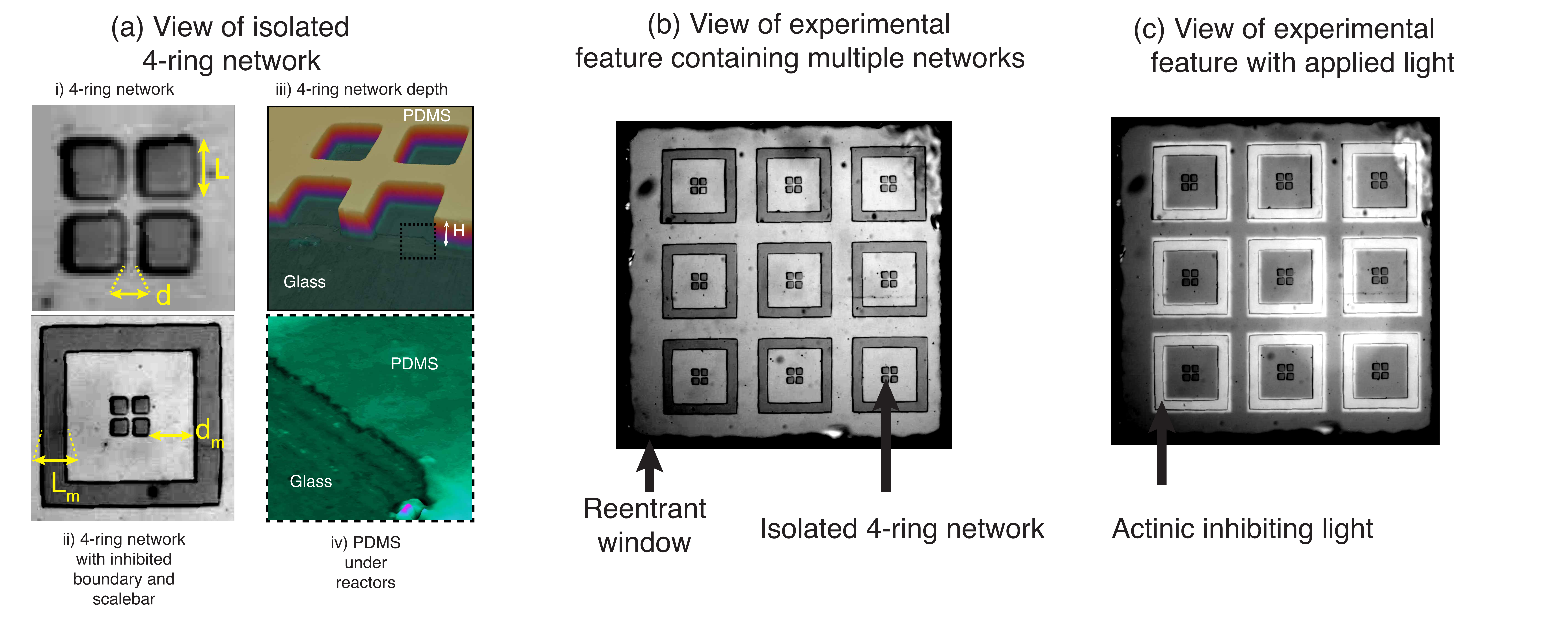}
\caption{
\textbf{(a)} i) A single network of four coupled BZ reactors. The side lengths, $L$,  of the reactors are $62$\si{\micro \metre}, side-to-side distances, $d$, between reactors are 26 \si{\micro \metre}.ii) The distance between a reactor and the inner surface of the inhibited boundary, $d_m$, was 130 or 190 \si{\micro \metre}. The thickness  of the inhibited boundary, $L_m$, was 100 \si{\micro \metre}. iii) 50X magnification profilometer image of cut microfluidic reactor network showing the out of plane height $H$ of the reactors is 30 \si{\micro \metre} and iv) 100X magnification profilometer image of cut microfluidic reactor network showing there is below 2\si{\micro \metre} of PDMS between the bottom of the reactors and the glass below it. 
\textbf{(b)} View of running experiment with nine parallelized  4-node ring networks running simultaneously. The reentrant window is a piece of glass used to seal the BZ into the lattice. The black region along the borders, outside the reentrant window, is filled with oscillatory BZ reaction \cite{Litschel2018}.
\textbf{(c)} View of same experiment except with actinic, inhibiting light selectively patterned onto the orders.}
\label{fig:viewexp}
\end{figure*}

\begin{figure*}[h!]
\includegraphics[scale=.125, angle=0]{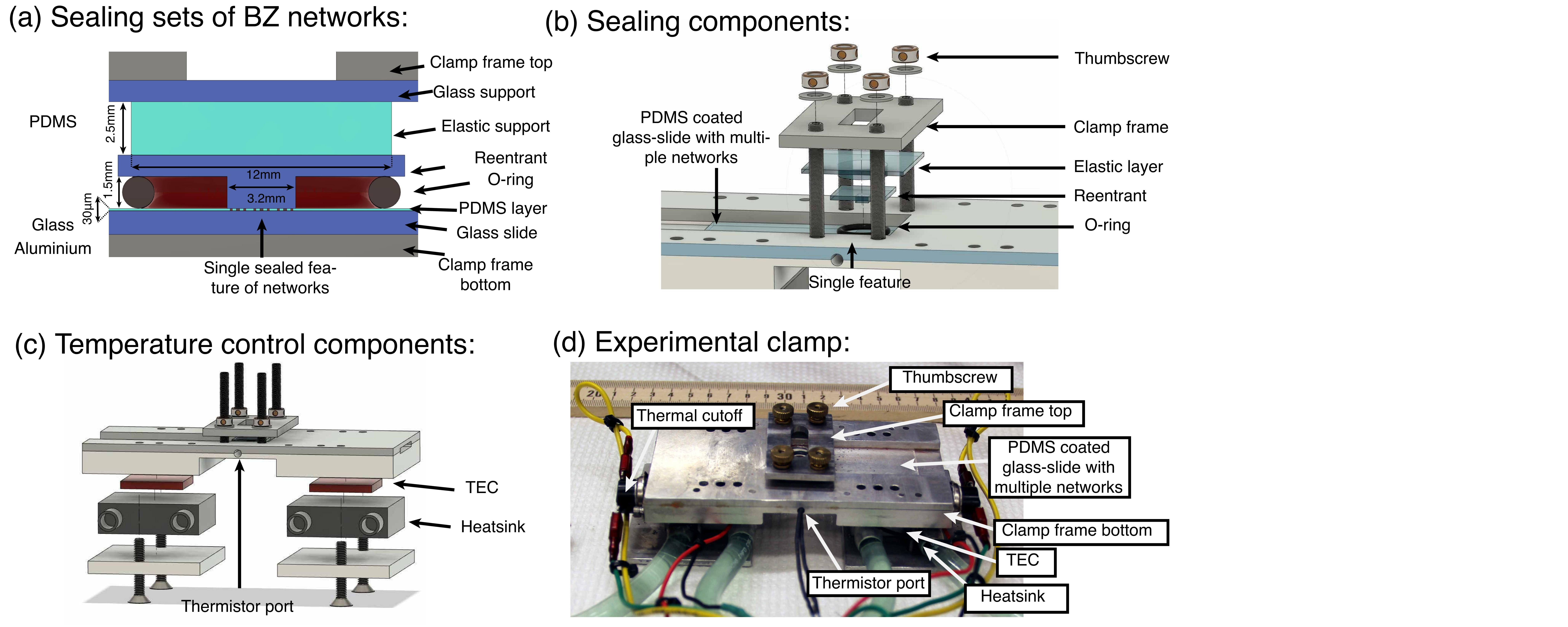}
\caption{Temperature controlled BZ clamp. 
\textbf{(a)} CAD diagram of sealing of a set, or feature, of BZ networks. The networks are sealed above and below by glass. A volume of extraneous BZ is also sealed, however is irrelevant to the dynamics of the network. Sealing is achieved by gradual adjustment of fine-threaded (6-80) thumbscrews. These apply force via a metal clamp frame, and softened by an 2.5mm thick layer of PDMS. 
\textbf{(b)} CAD diagram rendering of sealing components of the clamp. 
\textbf{(c)} CAD diagram of the temperature controlling elements of the clamp. Temperature measurement occurs through a thermistor. The clamp is heated and cooled using 2 thermoelectric cooler (TEC) peltier elements placed symmetrically about the sample. Feedback between them is performed using an Arduino's standard PID libraries. 
\textbf{(d)} Photograph of temperature-controlled clamp containing a sealed network. All components of the clamp are clearly visible except the TECs, elastic layer, reentrant window, and o-ring. Unlike in simplified diagrams (a),(b), and (c), three other necessary components of the clamp are pictured:  wiring, water circulating tubes connected to the heat sinks, and thermal cutoffs. The wiring connects the peltier TECs and thermistor to the controlling Arduino. The tubes circulate water through the heat sinks and cools the water down using fan radiators. The thermal cutoffs in series with the TECs short-circuit reversibly if the clamp become hot, above 50 \si{\celsius}, by some error. A meter stick is shown for reference (units of \si{\centi \metre}). }
\label{fig:tempcontrol}
\end{figure*}

\begin{figure*}[h!]
\includegraphics[scale=.1, angle=0]{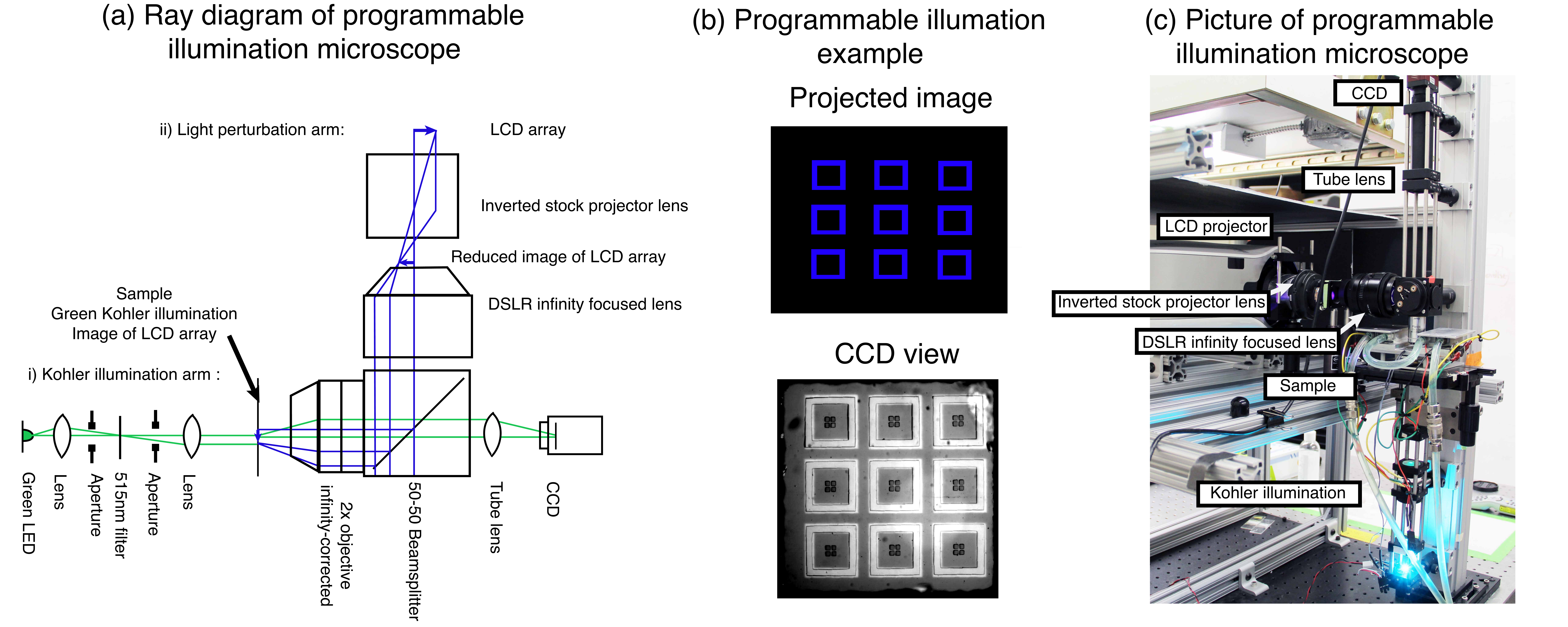}
\caption{
\textbf{(a)} Ray diagram of a both i) microscopy imaging of BZ networks and ii) their spatiotemporal perturbation with light. i) Imaging of the networks functions through 515nm green transmission illumination, ii) Spatiotemporal perturbation is achieved by imaging a commercial liquid-crystal display (LCD) projector's LCD array set to project pure blue light onto the sample. The blue light efficiently excites the photo-sensitive, oscillation-inhibiting catalyst. 
\textbf{(b)} Example of how spatiotemporal light patterning is achieved and its quality. Images are projected onto the sample by sending images to a VGA/HDMI port connected to the projector, an example is labeled 'Projected image'. This can be done through powerpoint, however we use MATLAB. This image then hits the sample, homogeneous and in crisp focus, shown in 'CCD view'. 
\textbf{(c)} Picture of full experimental apparatus, including a clamped, temperature-controlled sample. Overhead light shown in picture not on during experiments. 
}

\label{fig:optics}
\end{figure*}
. 

\begin{figure*}[h!]
\includegraphics[scale=.1, angle=0]{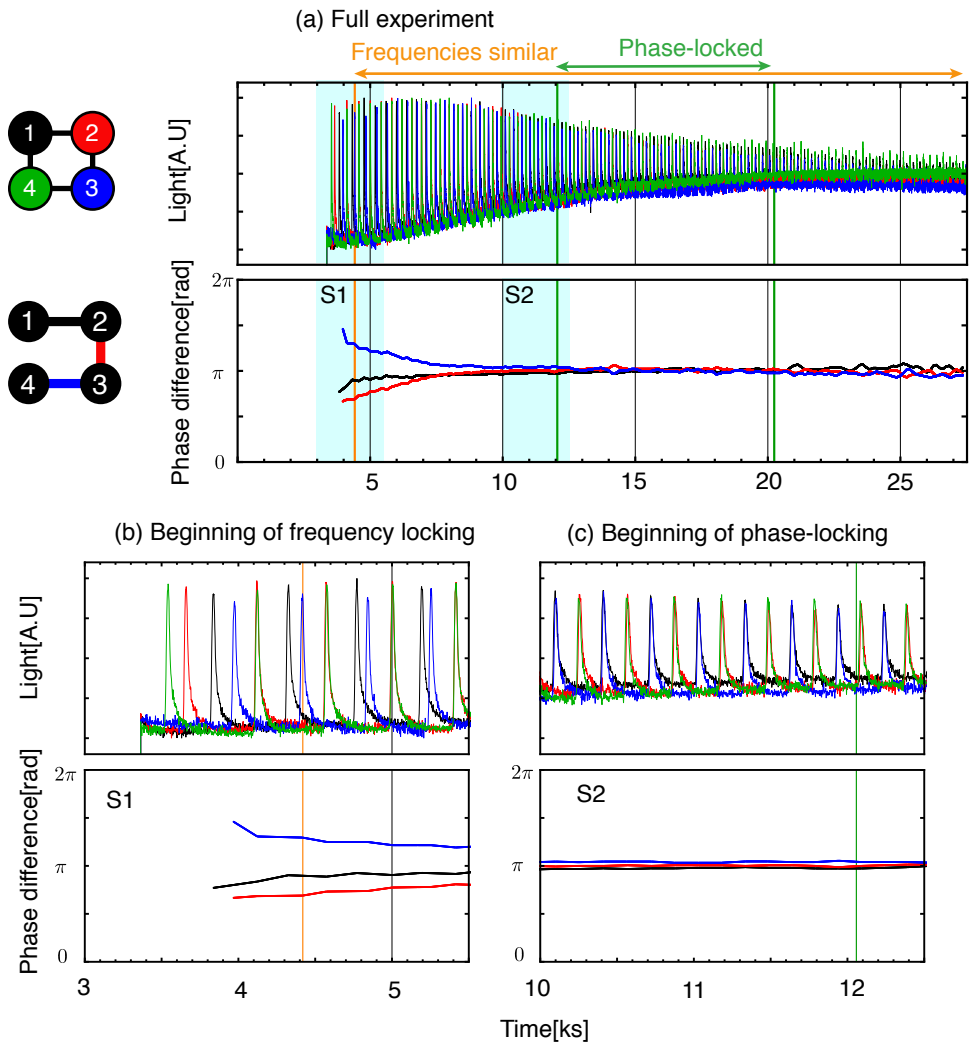}
\caption{
\textbf{(a)} View of a single 4 reactor network over time during an entire experiment. The top plot is of light transmitted through the reactors. It is normalized from the raw intensity measured as an average of a square of pixels inside a reactor [\fig \ref{fig:viewexp}(b)]. Specifically if the raw signal was $I(t)$, with a maximum value during the experiment was $I_{max} $ and a minimum at $I_{min} $, the normalized intensity was $(I(t)-I_{min})/(I_{max}-I_{min})$. As time goes on, the amplitude of the light peaks goes downwards, while the baseline increases. The lower panel tracks the phase differences between the reactors versus time. Eventually the reactors reach a phase-locked steady-state in which all reactors are firing antiphase their two neighbors. Since the phase differences between the reactors are never perfectly stationary, we use a specific threshold for when phase differences evolve slowly enough to be considered phase-locked in \apn{B}. At about 20,000\si{\second} the experiment ceases to be considered phase-locked as its phase differences become noisy. 
\textbf{(b)} An inset of early times in the experiment. Note that the frequencies and amplitudes of oscillation of the reactors are initially not the same. For example, the interval between the first two spikes of reactors 2(red) and 4(green) are different, and the amplitudes of 3(blue) and 4(green) are different. However, at about 4,400\si{\second} the reactors possess similar frequencies by the threshold in \apn{A} Protocol. 
\textbf{(c)} An inset of later times in the experiment. Note that the frequencies and amplitudes of oscillation of the reactors are very similar. At the green line marking 12,050\si{\second}, the threshold for phase-locking is met. 
}
\label{fig:exptraj}
\end{figure*}

\section{Explanation of best fit model}

\begin{figure*}[h!]
\includegraphics[scale=.15, angle=0]{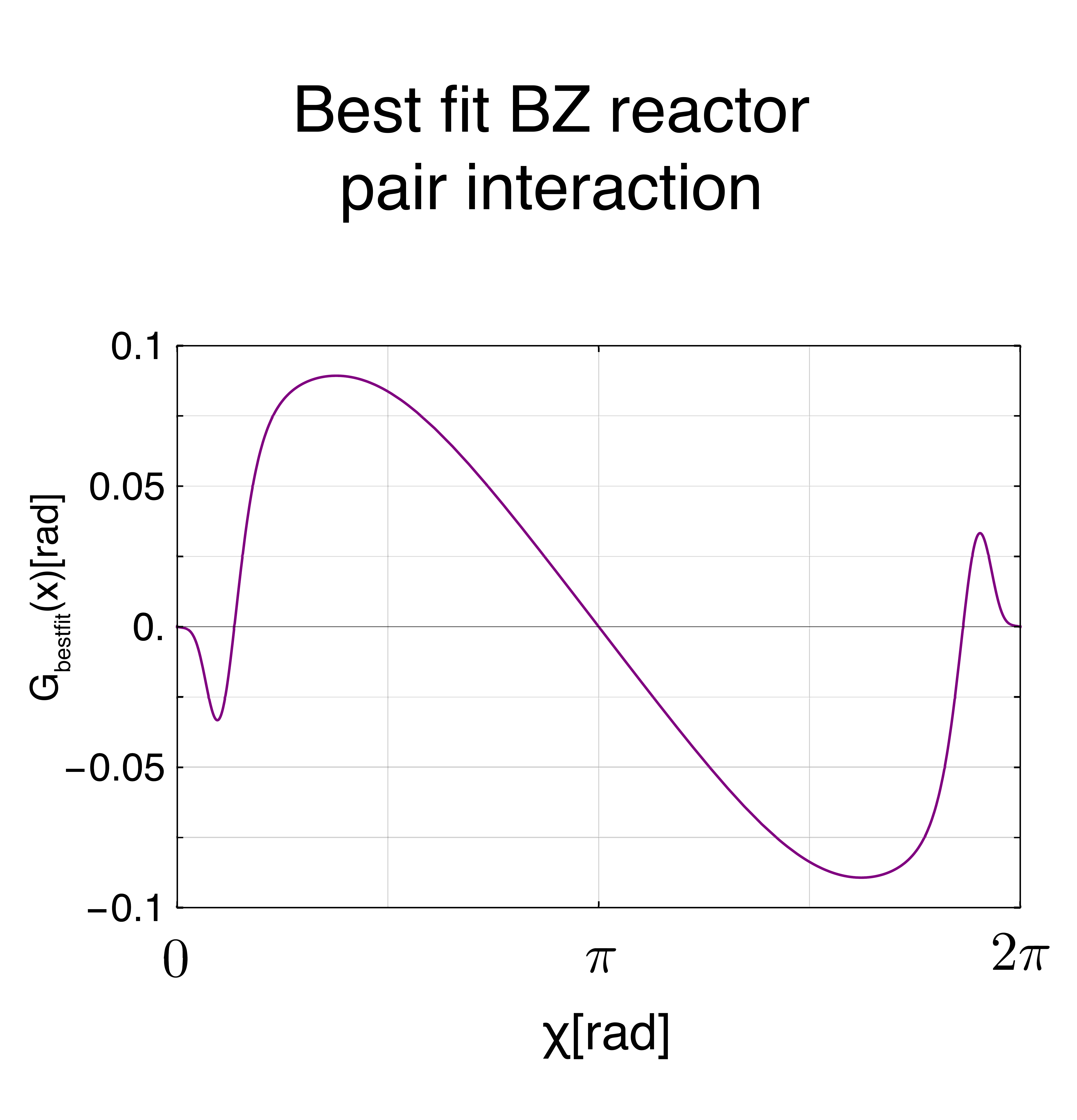}
\caption{The pairwise interaction function between two BZ reactors coupled by the best fit interaction function: $\frac{d}{dt} \theta_{21} = k G(\theta_{21})=k\left[H(-\theta_{21})-H(\theta_{21})\right]$. We can see that two negative slope zero crossings occur at $\theta_{21}=0,\pi$ - thus the best fit model has both in-phase and antiphase as stable attractors. }
\label{fig:Hfunc}
\end{figure*}

\subsection{Fitting to experiments/choice of free parameters}

\par The best fit model corresponds to substantially weaker than expected diffusive coupling. The diffusive coupling rate predicted by quasi-static diffusion between rectangular reactors was previously derived by Norton et al. \cite{Norton2019}: 
\begin{equation}
    k_{ideal} =\frac{PD}{Ld}
    \label{eqn:couplingrate}
\end{equation} where $P$ is the partition coefficient, $D$ the diffusion coefficient, $L$ the side length of the reactors, and $d$ is the distance between reactors \cite{tompkins_testing_2014,Norton2019}. In our case $D=10^{-9}$[\si{\metre \squared \per \second}], $P=2.5$[1], $L=62$[\si{\micro \metre}], and $d=26$[\si{\micro \metre}] thus the idealized coupling rate is $k_{ideal}=2.$[\si{\per \second}]. We see the best fit value of $k$ (manuscript Table V) is two orders of magnitude below $k_{ideal}$. In past works with emulsion droplets the best fit value of $k$ has been closer to one order of magnitude lower than expected \cite{tompkins_testing_2014,li_combined_2014,Norton2019}. Relative these past experiments with emulsion droplets in glass capillaries, densely packed emulsion droplets, and droplets in silicon chips, the experiments performed here had a much larger amount of oil-phase in the form of PDMS surrounding each 4 ring network \fig \ref{fig:viewexp}. As \ce{Br_2} is known to partition into and react with oil and PDMS, more PDMS would cause a reduction in inter-reactor coupling. We hypothesize this is the cause of this discrepancy.

\par Since the coupling rate equation \eqn \ref{eqn:couplingrate} is composed of known parameters $L$ and $d$, precisely prescribed by the photolithographic process when the chip is fabricated, and the product of the two less certainly known quantities $P$ and $D$, we can consider these best fits as a measurement of the product $PD$. Given the best fit $k=1.8\times 10^{-2}$[\si{\per \second}], $PD = 29.0 $[\si{\micro \metre \squared \per \second}].
 
\par Further, the best fit model corresponds to only slightly heterogeneous reactors, naively making it a near ideally symmetric network. To gain a sense for how heterogeneous the system is, we calculate the percent coefficient of variation in intrinsic frequencies of our best fit model. We consider a mean intrinsic frequency $\omega_o=\frac{2\pi}{300}$[\si{\radian \per \second}]. We can then consider the probability of an intrinsic frequency about the mean using the probability density function(PDF)  $\rho(\Delta \omega_{ij})=\rho(\omega_i-\omega_o)$ of our best fitted intrinsic frequency differences, manuscript \fig 6(c). Noting that the distribution of intrinsic frequencies $\rho(\omega_i)$ is then the same Laplacian PDF as $\rho(\omega_i-\omega_o)$, just shifted to be about $\omega_o$, we can see the variance $\sigma^2$ of intrinsic frequencies is the same as the variance of the PDF $\rho(\Delta \omega_{ij})$. We then explicitly calculate the variance of a Laplacian PDF, as it given by $\sigma^2 = 2b^2$ where $b$ is the rate parameter, so the standard deviation is $\sigma = \sqrt{2} b = \sqrt{2}\times2\pi7 \times 10^{-5}$\si{\radian \per \second}(manuscript Table V). Using this and the mean value $\omega_o$, we can compute a percent coefficient of variation $100*\frac{\sigma}{\omega_o}=10^{2}\frac{\sqrt{2} \times7\times 10^{-5}}{(3\times 10^2)^{-1}}=\sqrt{2}\times2.1=2.9[\%]$. This result from best fitting agrees with independent measurements of isolated, uncoupled reactors and shows the reactors are very similar in frequency in absolute terms. 

\par Yet, we can see this small degree of heterogeneity is sufficient to induce deviations from perfectly symmetric H/K derived results. How oscillators respond to heterogeneity doesn't just depend on  coupling strength $k$, it also depends on strength of the interaction function $H$. Letting $H_{max}$ be the maximum absolute value of $H$, in a pair of oscillators there is a critical nondimensionalized intrinsic frequency   $|\frac{\Delta \omega_{12}}{k H_{max}}|= \Delta \omega' \in [1,2]$ at which no phase-locked steady-state can exist past, initiating phase slipping. As the pair is symmetric when $\Delta \omega_{12} = 0$, the initiation of phase-slipping at finite $\Delta \omega_{12}$ is a point where the system is far from the fully symmetric limit. Although we are studying a larger network, so $\Delta \omega'>2$ no longer rigorously requires the network is phase slipping, the distribution of fit  $\Delta \omega'$  in manuscript \fig 6(c) allows us to estimate how far experiments are from being dynamically ideally symmetric. We see from this plot there are experiments which sample the tail of the exponential distribution of $|\frac{\Delta \omega_{ij}}{k H_{max}}|$ and are far from truly symmetric dynamics close to $|\frac{\Delta \omega_{ij}}{k H_{max}}|=1$. Thus, although the reactors possess very similar intrinsic frequencies,  the interaction function $H$ for our coupled BZ microreactors is sufficiently weak that experiments are dynamically not ideally symmetric. As a result, we can understand why experimental steady-states were at times distant from their ideal, H/K predicted values in manuscript \fig 5(a) and in \fig 6(a)(b), in spite of the reactors' small frequency dispersion.

\par The necessity of excitatory coupling is due to observation of in-phase synchronization during experiments. Without excitatory coupling, $k_e=0$, best fits to experiments with initial conditions that converged to Pronk and Bound were inaccurate as they would always converge to Trot. Although with no excitatory coupling, $k_e=0$, the phase model has Pronk and Bound as attractors, the basins of attraction of Pronk and Bound are much smaller than experimentally observed. Allowing finite excitatory coupling, specifically $k_e=.05$, achieved excellent fits to experimental data. 

\par Coupling across the diagonal was observed to have no impact. Surprisingly, $f$ at nearly any value from 0, a perfect ring with no cross-talk, to 1, a fully connected network, did not have a strong impact on the error of the fitting. Investigating the Pronk, Bound, Rotary Gallop, and Trot states, we found the stabilities of these states are only very weakly altered by the value of $f$. For this reason, $f=0$ was selected. 

\subsection{Best fit model:All steady-states}
\par
\begin{figure*}[h!]
\centering
\includegraphics[scale=.75, angle=0]{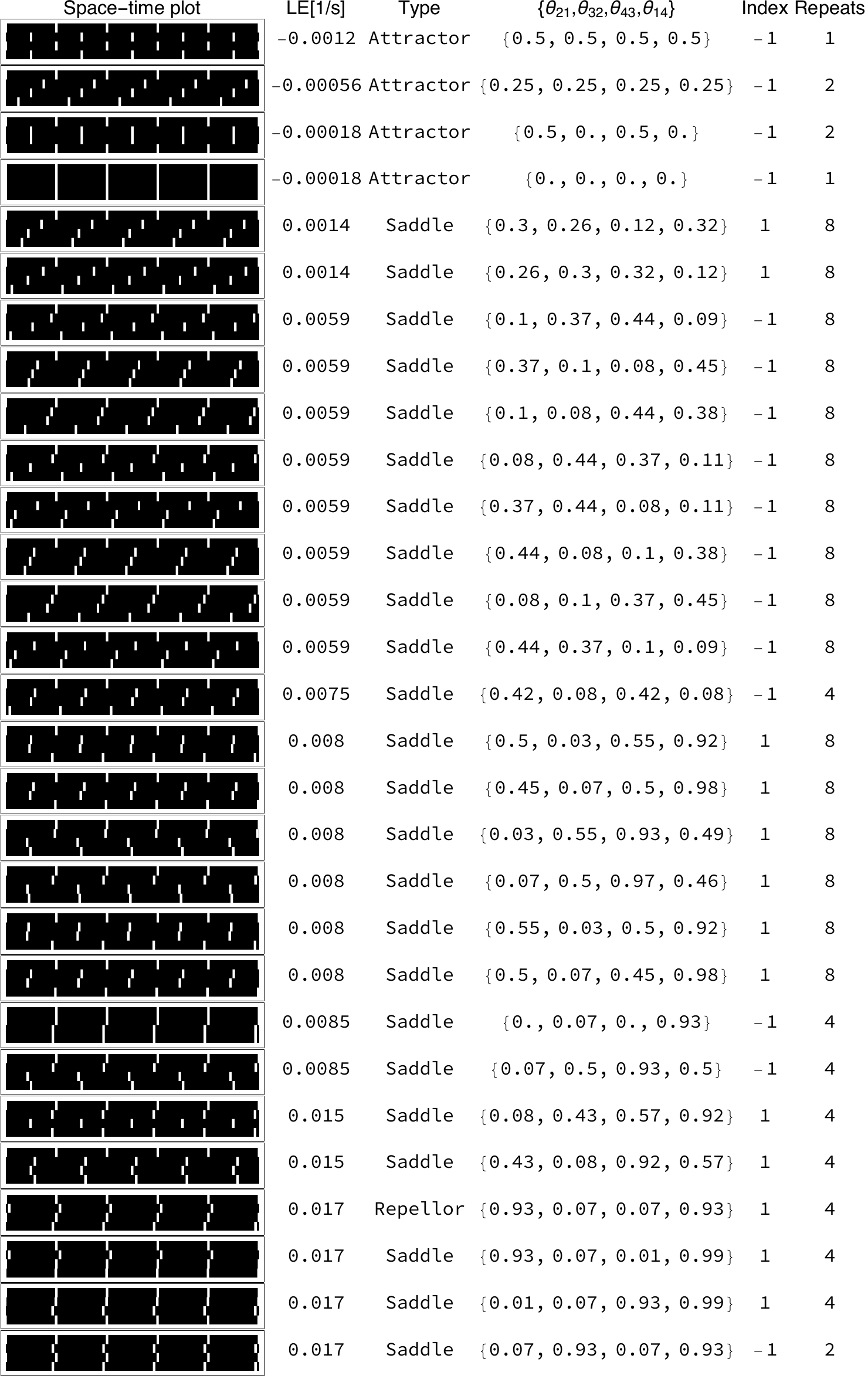}
\caption{A list of numerically discovered symmetrically unique  steady-states of the best fit model of the 4 ring network without heterogeneity.  For each state we computed: (i) An example of the space-time plot of the state, (ii) The Lyapunov exponent of the steady-state with units [\si{\per \second}], indicating the exponential rate at which nearby trajectory collapse to, if $\lambda<0$, or diverge from, if $\lambda>0$ , the steady-state, (iii) whether or not the state is an Attractor, Repellor, or a Saddle, (iv) the phase differences of the steady-state divided by $2\pi$, i.e. as fractions of a period (v) the topological index of the steady-state and (vi) the number of distinct states in state-space equivalent the unique state on the left. To be precise the number of distinct states is equal to the number of 4-tuples of phase differences $(\theta_{21},\theta_{32},\theta_{43},\theta_{14})$ generated by operating on the state on the left by every element of the symmetry group $D_4$ and then  modulo $2\pi$.
}
\label{fig:allsteadys}
\end{figure*}

\begin{figure*}[h!]
\centering
\includegraphics[scale=.2, angle=0]{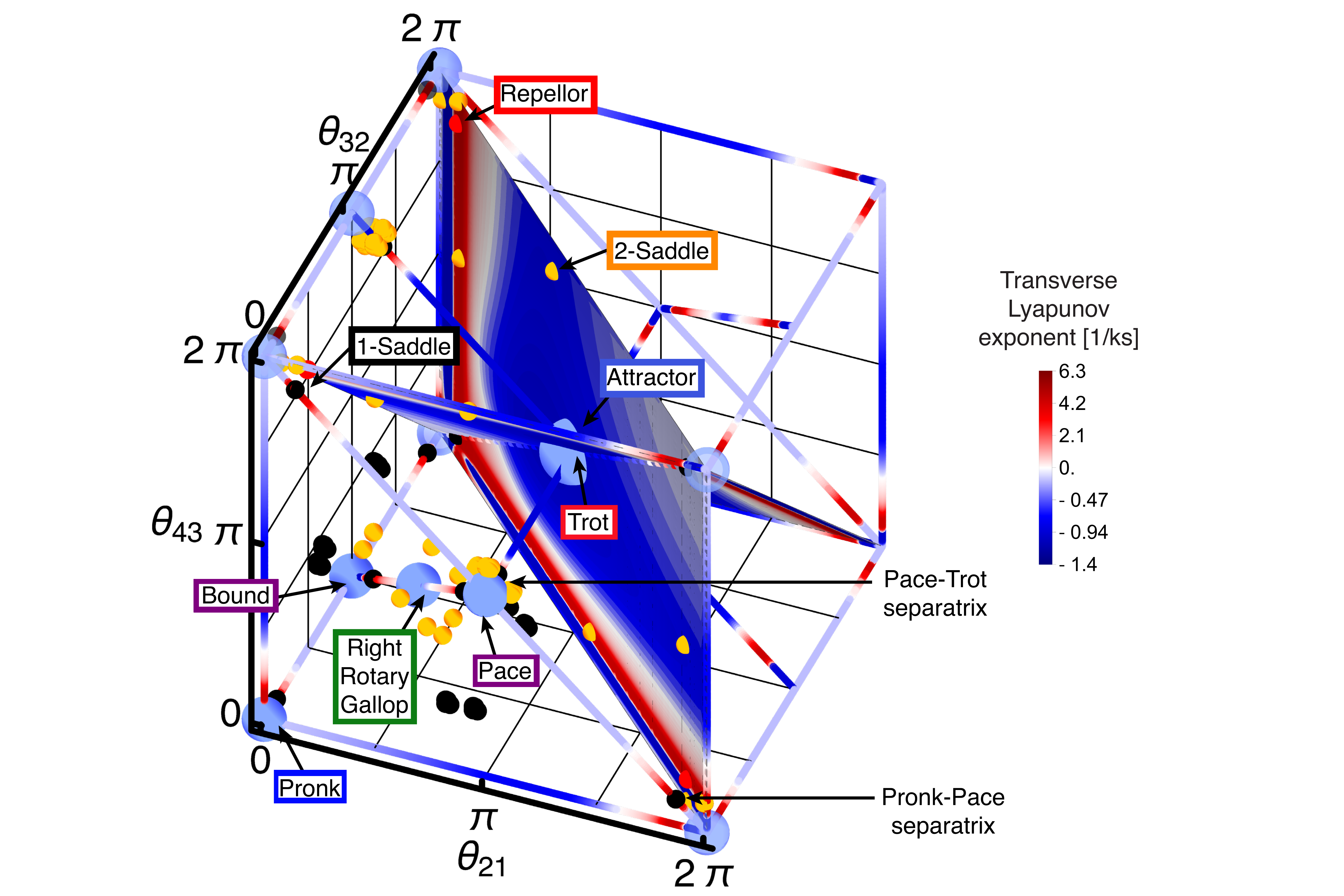}
\caption{ All steady-states of best fit model without heterogeneity which are in the half state-space $\theta_{32}>2\pi-\theta_{21}$, colored by type: Blue are Attractors with index=-1, Orange are saddles with 2 attracting direction saddles and thus index=1, Black are saddles with 1 attracting direction and thus with index=-1, Red Repellors with index=1.  Points are faint if they are a periodic instances of another steady-state. The unstable steady-stats which form the separatrices between Pronk and Pace, and Pace and Trot are also marked. 
}
\label{fig:allsteadysphasespace}
\end{figure*}

 In addition to the analysis of the best fit model included in the manuscript, we attempted to find all its steady-states: $\bar{\theta}'$ such that $\bar{\Psi}(\bar{\theta}')=\bar{0}$. This was achieved through dividing state-space into a NxMxM grid and initializing Newton-Ralphson rootfinders on $\bar{\Psi}(\bar{\theta})$ with no heterogeneity, $\Delta \omega_{ij} = 0$,  at the sites of that grid. The result of many iterations of the rootfinding $\bar{\theta}'_n$ was considered a valid steady-state if $\bar{\Psi}(\bar{\theta}'_n).\bar{\Psi}(\bar{\theta}'_n)<$ \num{1e-16}. The identified steady-states are shown as space-time plots with their Lyapunov exponent and topological index in \fig \ref{fig:allsteadys} and as colored points in the state-space in \fig \ref{fig:allsteadysphasespace}. The Lyapunov exponents were calculated  as the maximum real eigenvalue of the semi-analytically computed the Jacobian of the dynamics [\eqn \ref{eqn:jac}] at a given steady-state, utilizing numerically computed derivatives of the interaction function, $H$, manuscript \fig 6(a).  The state-space contains 168 steady-states \fig \ref{fig:allsteadys}; 10 of which are attractors or repellors while the remaining 158 are all saddle steady-states.  We believe this method may have discovered all steady-states since all indices sum to 0, the Euler characteristic of the state-space. Regardless of whether we found all steady-states, it does show: (1) The unique stable steady-states are the H/K point invariant manifolds and (2) the state-space possess a wealth of unstable, saddle steady-states. This method was too slow to be used to efficiently calculate the steady-states of the model with heterogeneity. 

\section{Computation stability or transverse Lyapunov exponents of the invariant manifolds}
\subsection{Example: computing  transverse Lyapunov exponents of an invariant manifold}

\par Following the algorithm present in \apn{E} for the invariant manifold $(D^p_1,D^p_1)$: 
\begin{enumerate}
\item{ We choose the invariant manifold $(D^p_1,D^p_1)$. It is a 2D manifold in the 3D state-space of the 4 ring network.}
\item{The invariant manifold  $(D^p_1,D^p_1)$ is parametrized by $\bar{\theta}=(f_1,-f_1,f_2)$, manuscript Table II. The set of tangent vectors which span it are $T = \{ \sqrt{2}^{-1}(1,-1,0),(0,0,1) \}$}. The one vector normal, which we put in our set of possible normal vectors $N=\{ 2^{-\frac{1}{2}}(1,1,0) \}$. 
\item{We now compute the unitary transformation matrix $P$. In the case of the 4 ring's invariant manifold this is very simple:
$$P =\left[\bar{t}_1 \bar{t}_2  \bar{n}_1\right]^T = \frac{1}{\sqrt{2}} 
\begin{bmatrix} 
1  & -1 & 0\\
0 & 0& \sqrt{2}\\
1  & 1& 0\\
\end{bmatrix}$$}
\item{For our particular 4 ring network the Jacobian can be analytically computed:
$H'(x)=\frac{d H(\chi)}{d \chi}|_{\chi=x}$,  $J(\bar{\theta})=$
\begin{gather}
    -k  \Bigg[\begin{smallmatrix}
    \left[ H'(\theta_{21})+H'(-\theta_{21})+H'(\theta_{43}+\theta_{32}+\theta_{21})\right] & 
     \left[ -H'(\theta_{32})+H'(\theta_{43}+\theta_{32}+\theta_{21})\right] &
       H'(\theta_{43}+\theta_{32}+\theta_{21}) \\
     -H'(-\theta_{21}) & 
     \left[ H'(\theta_{32})+H'(-\theta_{32}) \right] &
       -H'(\theta_{43})\\
     H'(-(\theta_{43}+\theta_{32}+\theta_{21}))& 
     \left[ H'(-(\theta_{43}+\theta_{32}+\theta_{21})) -H'(-\theta_{32})\right] &
      \left[ H'(\theta_{43})+H'(-\theta_{43})+H'(-(\theta_{43}+\theta_{32}+\theta_{21}))\right]\\
    \end{smallmatrix}  \Bigg] \nonumber
    \\  +fk \begin{bmatrix}
    H'(\theta_{32}+\theta_{21}) & 
     \left[ -H'(\theta_{43}+\theta_{32})+H'(\theta_{32}+\theta_{21})\right] &
      -H'(\theta_{43}+\theta_{32})  \\
      H'(-(\theta_{32}+\theta_{21})) & 
     \left[ H'(-(\theta_{32}+\theta_{21}))+H'(\theta_{43}+\theta_{32}) \right] &
       H'(\theta_{43}+\theta_{32})\\
     -H'(-(\theta_{32}+\theta_{21}))& 
     \left[ H'(-(\theta_{43}+\theta_{32})) -H'(-(\theta_{32}+\theta_{21}))\right] &
      H'(-(\theta_{43}+\theta_{32})) \\
    \end{bmatrix}
    \label{eqn:jac}
    \end{gather}
    Note that since we numerically computed $H$, $H'$ is computed numerically as well. 
    }
    \item{
    We compute $J'=P.J.P^{-1}$. We let $f=0$ for simplicity. 
    $$J' = \begin{bmatrix}
    -2H'(-\theta_{21})-H'(\theta_{21}) & -2H'(\theta_{43}) & -H'(\theta_{43}) \\ 
    -\frac{1}{2}H'(\theta_{21})    & -2H'(-\theta_{43})-H'(\theta_{43}) & \frac{1}{2}(H'(\theta_{21})-2H'(-\theta_{43})) \\
    0 & 0 & -H'(\theta_{21})-H'(\theta_{43})
    \end{bmatrix}$$
    }
    \item{We extract the 'block' of dynamics corresponding to perturbations in the only normal direction
    $$\left[ -H'(\theta_{21})-H'(\theta_{43}) \right]$$ 
    }
    \item{In general, this extracted block will not be 1x1 and the max Transverse Lyapunov exponent will be its max real eigenvalue. Since the matrix is 1x1 the value itself is the MTLE. We conclude that a infinitesimally small perturbation normal the $(D^p_1,D^p_1)$ invariant manifold $\delta n$ will evolve according to the equation $$\frac{d}{dt}\delta n =\left[-H'(\theta_{21}(t))-H'(\theta_{43}(t)) \right] \delta n$$. 
    }
\end{enumerate}

\subsection{Maximum transverse lyapunov exponents of all weakly coupled 4 ring networks}
We now present the MTLE for a 4 ring network of phase oscillators for a generic phase model with any interaction function $H$. With the exception of the linear invariant H/K manifold $(Z_2,1)$, all of them are simple.

We used the algorithm proposed in the previous section on each invariant manifold to transform this Jacobian $J$ to identify the dynamics transverse to the manifold in $J'$. The result of these calculations are shown in Table \ref{table:transverseLE} is the block of $J'$ corresponding to components normal or transverse the each invariant manifold. 

\begin{center}
\begin{table}[h!]
\centering
\caption{Decoupled transverse linear dynamics of invariant manifolds universal to four-rings($f=0$). $H'(x) = \frac{d H(\chi)}{d \chi}|\chi=x$. Note that the max transverse Lyapunov exponent of an invariant manifold is the maximum real eigenvalue of the matrix in the case of the linear invariant manifolds and a scalar in the case of the planar invariant manifolds}
\setlength{\tabcolsep}{5pt} 
\renewcommand{\arraystretch}{1} 
\begin{tabular}{|K{10mm}|K{4.5mm}|K{80mm}|} 
\hline
H & K &Transformed Jacobian $J'$ normal dynamics block\\
\hline
\multicolumn{3}{|c|}{\text{\small Linear invariant manifolds:} } \\
\hline
$D^s_1$ & $D^s_1$ & $ -\begin{bmatrix}
2 H'(0) + H'(-\theta_{32}) & H'(-\theta_{32}) \\
 H'(\theta_{32}) & 2 H'(0) + H'(\theta_{32}) \\
\end{bmatrix}$ \text{ }\\
\hline
$D^s_1$ & $1$ &
$ -\begin{bmatrix}
2H'(\pi)+H'(-\theta_{32}) & H'(-\theta_{32}) \\ 
H'(\theta_{32}) & 2H'(\pi)+H'(\theta{32}) \\
\end{bmatrix}$  
\\ 
\hline
$Z_2$ & $1$ & 
$\begin{bmatrix}
J'(z2,1)_{11} & J'(z2,1)_{12} \\ 
J'(z2,1)_{21} & J'(z2,1)_{22}\\
\end{bmatrix}$  \\ 
\hline
\multicolumn{3}{|c|}{\text{\small Planar invariant manifolds:} } \\
\hline
$D^p_1$ & $D^p_1$ & $-H'(\theta_{21})-H'(\theta_{43})$ \\
\hline
\end{tabular}
\label{table:transverseLE} 
\end{table}
\end{center}
Terms for $(Z_2,1)$ invariant manifold:\newline
$J'(z2,1)_{11}=\frac{1}{3}(2H'(\pi+\theta_{21})+H'(-\pi+\theta_{21})-3H'(\theta_{21})-3H'(-\theta_{21})-H'(-\theta_{21}+\pi))-2H'(-\theta_{21}-\pi))$
\newline
$J'(z2,1)_{12}=-\frac{2}{3}(
-2H'(\pi-\theta_{21})-H'(-\pi+\theta_{21})
+H'(\pi-\theta_{21})+2H'(-\pi-\theta_{21})
)
$
\newline
$J'(z2,1)_{21}=\frac{1}{3}(
-2H'(\pi+\theta_{21})-H'(-\pi+\theta_{21})
+3H'(\theta_{21})
)$
\newline
$J'(z2,1)_{22}=-\frac{2}{3}(2H'(\pi+\theta_{21})+H'(-\pi+\theta_{21}))$


\subsection{Transverse stability of invariant manifolds: Kuramoto model compared to BZ phase model}
\par
\begin{figure}
\includegraphics[scale=.125]{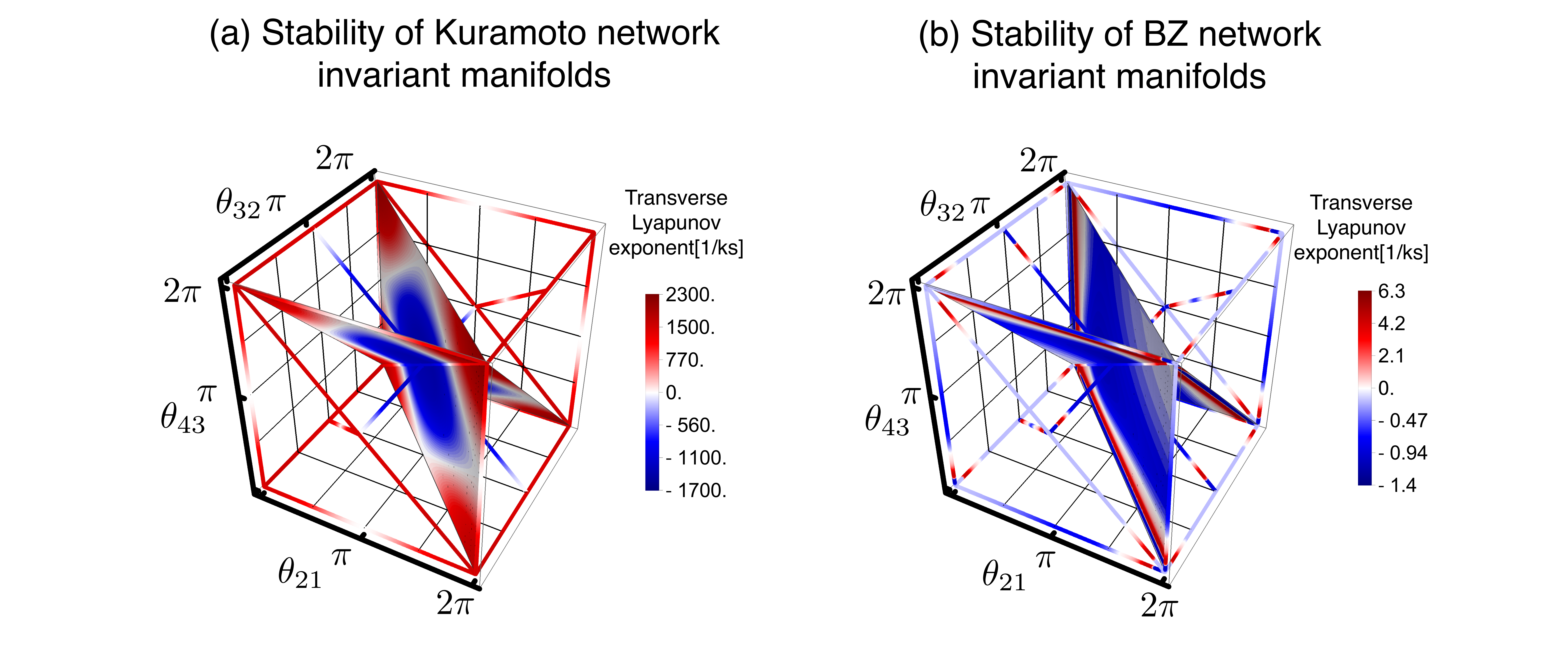}
\caption{ 
\textbf{(a)} Maximum transverse Lyapunov exponent of 4 ring of inhibitory Kuramoto oscillators invariant manifolds. The $(D^s_1,D^s_1)$ and $(Z_2,1)$  invariant manifolds are fully repelling. The $(D^p_1,D^p_1)$ invariant manifolds are half repelling and half attracting.  The $(D^s_1,1)$ manifold is largely attracting.  
\textbf{(b)} Maximum transverse Lyapunov exponent of best fit model of 4 ring of inhibitory BZ oscillators. All invariant manifolds are largely attracting, except $(Z_2,1)$ 
} 
  \label{fig:contractKvsBZ}
\end{figure}

 Having both parametrized guaranteed invariant manifolds (manuscript Table II) and provided firm expressions for the MTLE of 4 ring networks' invariant manifolds (Table \ref{table:transverseLE}), we compare the invariant manifolds of 4 ring Kuramoto model networks and those of the best fit model of coupled BZ reaction-diffusion network. In previous works some similarity was seen between the simplified Kuramoto model and inhibitory coupled BZ reactors in hexagonal lattices \cite{Li2015,giver_phase_2011}. The MTLE for networks of 4 nodes with perfect dihedral 4 symmetry (A) Kuramoto 4 ring, model with $H(\chi)=-\sin(\chi),f=0 \text{ and } k=1$,  and (B) BZ best fit model [manuscript Table V] are shown as heatmaps applied to H/K invariant manifolds in \fig \ref{fig:contractKvsBZ}. 
\par We observe that the two 4 node networks are qualitatively different. The Kuramoto model's invariant manifolds are exclusively repelling except of $(D^s_1,1)$ and $(D^p_1,D^p_1)$ about Trot. The best fit BZ model's invariant manifolds are almost exclusively attracting, with the exception of $(Z_2,1)$, which is half attracting and half repelling. We conclude that, although superficially the two interactions are similar, they are unlike since the transient dynamics of the BZ models will often form symmetric clusters on H/K derived invariant manifolds before reaching steady-state, while the Kuramoto model will not.  

\section{Euler characteristic of a 3-Torus}
The Euler characteristic of a solid Polyhedra is given by:
$$\chi = V-E+F-C$$
which depends on integer number of vertices $V$, edges $E$, faces $F$, and number of 3D objects $C$. The 3-Torus is a cartesian cube with periodic boundary conditions. Thus it is a cubic polyhedra, which very few unique vertices, edges, and faces.
\par Vertices: There are 8 vertices of a normal cube. If we consider any vertex at the edge of the periodic cube, we see they are related to one another by periodic boundary conditions. $V=1$. 
\par Edges: There are 12 edges of a normal cube. Each edge of the periodic cube is repeated 3 times. For example the edge $((0,0,0),(0,0,2 \pi))$ is equal to $((2\pi,0,0),(2 \pi,0,2 \pi))$,$((0,0,0),(0,2\pi,2 \pi))$, and $((2\pi,2\pi,0),(2\pi,2\pi,2 \pi))$. There are thus 3 unique edges $E=3$
\par Faces: There are 6 faces to a normal cube. In the periodic cube each face is repeated twice. $F=3$
Thus the Euler characteristic is 0:$\chi = 1-3+3-1=0$

\bibliography{FourRingPaperAV1.bib}